\documentclass[11pt]{article} 
\pdfoutput=1 

\usepackage{jheppub} 
\usepackage{tikz}
\usepackage{tikz-feynman}
\usepackage[utf8]{inputenc}
\usepackage[english]{babel}
\usepackage{amsmath,amssymb,amsbsy,amstext,amsthm,booktabs,simplewick,exscale,relsize,slashed,graphicx,amsfonts,upgreek, xcolor,subcaption}
\usepackage{multirow,color,dcolumn,bm,enumerate}
\usepackage{hyperref}
\usepackage{feynmp-auto,axodraw2,tikz}
\usepackage{ulem}
\usepackage{comment}
\usepackage{amsmath}
\usepackage{ytableau}
\usepackage{braket}
\usepackage{booktabs}
\DeclareUnicodeCharacter{2212}{-}

\usepackage{hyperref}
\usepackage[capitalise,noabbrev,nameinlink]{cleveref}

\usepackage{mathtools}

\title{Energy-Enhanced Dimension Eight SMEFT Effects in VBF Higgs production}
\author[a,b]{Benoît Assi,}
\author[c]{Adam Martin}
\affiliation[a]{Particle Theory Department, Fermilab, Batavia, IL, 60510,  USA}
\affiliation[b]{Department of Physics, University of Cincinnati, Cincinnati, OH, 45221, USA}
\affiliation[c]{Department of Physics, University of Notre Dame,
  South Bend, IN, 46556 USA}

\emailAdd{amarti41@nd.edu, bassi@fnal.gov}

\abstract{We study Higgs boson production via vector boson fusion at the LHC, focusing on the process $pp \to H + jj$ and capturing the leading energy-enhanced contributions within the Standard Model Effective Field Theory (SMEFT) up to order $1/\Lambda^4$. Employing energy-scaling arguments, we predict the magnitude of each higher-dimensional operator's contribution. Utilizing the geometric formulation of SMEFT, our analysis incorporates dimension-eight operators not previously considered. We find that the kinematics of vector boson fusion—characterized by two highly forward jets—tend to suppress contributions from higher-dimensional operators, requiring a lower scale $\Lambda$ for SMEFT effects to become observable. This suggests that the SMEFT remains valid for lower $\Lambda$ than expected. Combined with the fact that LEP constrains the dimension-six operators with the most considerable impact on vector boson fusion, a regime exists where dimension-eight operators can have significant effects. In many cases, these dimension-eight operators also influence associated production processes like $pp \to H\, V(jj)$, though differences in analysis cuts and kinematics mean this is not always the case. Our findings provide insights that could refine the search for SMEFT signals in collider experiments.
}

\begin{document}
\maketitle
\allowdisplaybreaks[1]
\setcounter{page}{2}
\newpage
\section{Introduction}\label{sec:intro}

Vector boson fusion (VBF) is an essential process for Higgs boson studies at the Large Hadron Collider (LHC)~\cite{Dittmaier:2012vm,LHCHiggsCrossSectionWorkingGroup:2011wcg,LHCHiggsCrossSectionWorkingGroup:2013rie,LHCHiggsCrossSectionWorkingGroup:2016ypw}. As the second-largest Higgs production mechanism after gluon fusion, VBF is uniquely characterized by its dependence solely on electroweak interactions, which distinguishes it from other production modes that also involve strong force couplings. This purely electroweak nature makes VBF particularly sensitive to the Higgs boson's properties and any potential deviations from the Standard Model (SM) predictions. The process's clean signature and electroweak purity allow for precise measurements and provide a clear path for probing the Higgs sector and searching for new physics.

The experimental signature of VBF Higgs production is distinctive, marked by two jets that are observed in opposite hemispheres of the detector. These jets have a large invariant mass, denoted as \(m_{j_1j_2}\), and are separated by a large pseudorapidity gap, \(|\Delta \eta_{j_1j_2}|\). This configuration leads to a unique topology that helps to reduce backgrounds from other processes, enhancing the ability to isolate VBF events from the plethora of interactions within the LHC. The specific characteristics of these jets—not only their mass and separation but also their minimal hadronic activity between them—serve as key indicators in identifying and analyzing VBF Higgs production, making it a critical focus for precision SM physics as well as beyond the SM (BSM) searches.

One powerful approach to probe beyond the SM in this context is the application of Effective Field Theory (EFT), which provides a framework to incorporate potential new physics effects systematically. The Standard Model Effective Field Theory (SMEFT) is now the standard framework for this task~\cite{Araz:2020zyh,Buchmuller:1985jz, Giudice:2007fh, Grzadkowski:2010es, Gupta:2011be,Gupta:2012mi,Banerjee:2012xc, Gupta:2012fy, Banerjee:2013apa, Gupta:2013zza,Elias-Miro:2013eta, Contino:2013kra, Falkowski:2014tna,Englert:2014cva, Gupta:2014rxa, Amar:2014fpa, Buschmann:2014sia, Craig:2014una, Ellis:2014dva, Ellis:2014jta, Banerjee:2015bla, Englert:2015hrx, Ghosh:2015gpa, Degrande:2016dqg,Cohen:2016bsd, Ge:2016zro, Contino:2016jqw, Biekotter:2016ecg, deBlas:2016ojx, Denizli:2017pyu, Barklow:2017suo, Brivio:2017vri, Barklow:2017awn, Khanpour:2017cfq, Englert:2017aqb, panico,Franceschini:2017xkh, banerjee1, Grojean:2018dqj,Biekotter:2018rhp, Goncalves:2018ptp,Gomez-Ambrosio:2018pnl, Freitas:2019hbk,Banerjee:2019pks, Banerjee:2019twi, Biekotter:2020flu}. In SMEFT, the SM Lagrangian is extended by adding a series of higher-dimensional operators, which represent potential deviations from the SM due to new physics at a scale \(\Lambda\) not directly accessible by the LHC. The SMEFT Lagrangian is given by:
\[
\mathcal{L}_{\rm SMEFT} = \mathcal{L}_{\rm SM} + \sum_{i,j} \frac{c_j}{\Lambda^i} \mathcal{O}_j^{(4+i)}
\]
where \(i\) indexes the dimensions greater than four, while \(c_i\) and \(\mathcal{O}_i\) indexes all possible Wilson coefficients and corresponding operators of a given mass dimension. We will often refer to the operators by their corresponding Wilson coefficients in what follows. We will ignore operators of odd mass dimensions as these violate baryon ($B$) and lepton ($L$) number conservation~\cite{DAmbrosio:2002vsn}.  The SMEFT allows for examining how these higher-dimensional operators might alter both the production cross-section and the decay characteristics of the Higgs boson, potentially revealing new dynamical effects from physics well above the electroweak scale. Assuming $B$ and $L$ are preserved, the leading contributions for LHC physics often arise at dimension six~\cite{Almeida:2021asy,Brivio:2021alv,Ellis:2020unq}, corresponding to corrections of order \(1/\Lambda^2\). 

This work aims to study VBF Higgs production to $\mathcal{O}(1/\Lambda^4)$, as has been achieved recently in various other important LHC processes~\cite{Hays:2018zze, Corbett:2021jox,Corbett:2023qtg,Martin:2023fad,Boughezal:2021tih,Corbett:2021cil,Alioli:2020kez,Boughezal:2022nof,Kim:2022amu, Allwicher:2022gkm,Dawson:2021xei,Degrande:2023iob,Corbett:2021eux,Martin:2023tvi}. 
We push to dimension eight mainly due to the so-called ``inverse problem", where the degeneracy between different UV completions of the SM at dimension six remains too broad. This degeneracy is significantly reduced at dimension eight, making it easier to distinguish between different possibilities for new physics~\cite{Zhang:2021eeo}.
Previous authors~\cite{Araz:2020zyh} identified that certain dimension-six operators, which can be represented as \( (\psi^\dag\bar{\sigma}^\mu \psi)(i H^\dag \overleftrightarrow{D}_\mu H) \) in the Warsaw basis~\cite{Grzadkowski:2010es}, result in the largest energy enhancement in VBF. We can understand the impact of these operators by examining the subprocess $q\,V \to q\,H$. Once this subprocess is sewn into a $qqV$ vertex, it contributes to VBF; however, as $q\,V \to q\,H$ is a $2 \to 2$ process, it is simpler to analyze analytically. The SM contributions to the $q\,V \to q\,H$ amplitude are $\sim v^2/\sqrt{\hat t}$, where we use $\hat t$ as a proxy for the energy-squared flowing through the diagram. The key aspect of the \( (\psi^\dag\bar{\sigma}^\mu \psi)(i H^\dag \overleftrightarrow{D}_\mu H) \) operators is that they generate a four-point $\psi^2VH$ contact vertex which enters the $q\,V \to q\,H$ amplitude $\sim v\sqrt{\hat{t}}/\Lambda^2$. Combining these two pieces of the amplitude and squaring, we see the interference piece is enhanced (relative to the SM) at large $\hat t$,
\begin{align}
\frac{|A_{\rm SM}A^*_{\psi^2H^2D}|}{|A_{\rm SM}|^2}\Big|_{qV \to q\,H} \sim \frac{\hat t}{\Lambda^2}.
\label{eq:basicE}
\end{align}
Repeating the exercise with other dimension six operators, we always find a smaller ratio. We further note that operators involving tensor structures, in particular, \(H^2X^2\) operators, might have unique interference patterns depending on the polarization states of the particles involved, leading to potentially significant effects that warrant further detailed studies of their impact on polarization observables.

Besides affecting the $qqVH$ vertex, operators of the form \( (\psi^\dag\bar{\sigma}^\mu \psi)(i H^\dag \overleftrightarrow{D}_\mu H) \) also contribute to $qqV$ vertices when both Higgs fields are set to their vacuum expectation value. Precision electroweak (EW) observables highly constrain these contributions from LEP. Reference~\cite{Araz:2020zyh} highlighted the sensitivity of VBF to these constrained operators as an opportunity to leverage the energy enhancement in Eq.~\eqref{eq:basicE} to probe small Wilson coefficients, potentially improving upon LEP constraints.

Expanding VBF within SMEFT to order $\mathcal{O}(1/\Lambda^4)$ reveals several new and interesting features:

\begin{itemize}

\item Hierarchies in the Wilson coefficients can disrupt the SMEFT power counting, allowing formally higher-order terms in $\mathcal{O}(1/\Lambda)$ to have significant effects, especially in the tails of kinematic distributions. In this context, higher-order terms refer to those of $\mathcal{O}(1/\Lambda^4)$, which can arise from squared dimension-six operators—meaning both the literal square of each SMEFT operator and the interference among different dimension-six terms—or from the interference between dimension-eight operators and the SM. When forming the analogous ratio to Eq.~\eqref{eq:basicE} that includes $\mathcal{O}(1/\Lambda^4)$ terms, the numerator must carry four powers of energy. This energy dependence can involve any combination of the Higgs vacuum expectation value $v$ and the process energy scale $\sqrt{\hat{t}}$. Following Ref.~\cite{Araz:2020zyh}, we focus on the regime where $\sqrt{\hat{t}} > v$, so operators with more powers of $\sqrt{\hat{t}}$ are considered more "energy enhanced".

\item While the classification of EFT effects using $\sqrt{\hat{t}}/v$ sufficed at order $\mathcal{O}(1/\Lambda^2)$, we must exercise greater caution at order $\mathcal{O}(1/\Lambda^4)$ due to the complexity of VBF as a $2 \to 3$ process. VBF is characterized by $\hat{t} \ll \hat{s}$, but unlike simpler $2 \to 2$ processes, distinguishing between EFT effects proportional to $\hat{t}/\Lambda^2$ and those proportional to $\hat{s}/\Lambda^2$ is more subtle. At dimension six, SMEFT effects in VBF predominantly scale with $\hat{t}$, while at dimension eight, contributions involving both $\hat{t}$ and $\hat{s}$ emerge. This interplay requires careful analysis, as the energy dependence in a $2 \to 3$ process like VBF does not directly translate from the simpler $2 \to 2$ cases.

\item At order $\mathcal{O}(1/\Lambda^4)$, there are SMEFT five-particle contact vertices that contribute to VBF. These vertices originate from operators of the form $H^2\psi^4$ and are attractive for several reasons. First, they lead to VBF amplitudes without internal propagators, suggesting an enhancement relative to the SM at high energies. Second, while all the other operators we consider also contribute to crossed processes like $pp \to h\, V(\ell\ell)$ or $pp \to h\, V(jj)$, we expect that these particular operators do not, due to the specific kinematic configurations and selection cuts inherent to VBF. 

\end{itemize}

This work is structured as follows. In Section \ref{sec:dim8ops}, we introduce the theoretical background and significance of higher-dimensional operators with energy-enhanced effects in VBF. Section \ref{sec:geobasis} elaborates on the geometric approach to the SMEFT and the selection of compatible operator bases. The analysis of energy-enhanced contributions to VBF processes is presented in Section \ref{sec:SMEFTcont}. Section \ref{sec:pol} discusses specific resonant operators and their characteristics. In Section \ref{sec:distributions}, we present the expected distributions of observables affected by these operators. Section \ref{sec:associated} explores crossed processes, particularly the associated production in \(pp \to VH\). Finally, Section \ref{sec:conclude} concludes the paper with a summary of our findings.

\section{Energy-enhanced effects of dimension eight operators in VBF}\label{sec:dim8ops}

 This section examines how dimension-eight operators contribute to VBF processes, focusing on their energy-enhanced effects at order $1/\Lambda^4$. We identify three key issues that arise when including these higher-order terms: first, hierarchies among Wilson coefficients can disrupt the usual SMEFT power counting, enhancing the effects of dimension-eight operators relative to dimension-six ones; second, subtleties in energy scaling become significant due to the specific kinematics of $2 \to 3$ processes like VBF, requiring careful analysis of how energy enters the SMEFT contributions; and third, the introduction of new five-particle contact vertices at dimension eight leads to additional contributions to VBF amplitudes.

Within the framework of the SMEFT, the amplitude squared for a specific process can be described by the following equation
\begin{align}
|\mathcal A|^2 = |A_{\rm SM}|^2 \Big\{ 1 + \frac{2 \text{Re}(A^*_{\rm SM}A_6)}{\Lambda^2 |A_{\rm SM}|^2} + \frac{1}{\Lambda^4}\Big(\frac{|A_6|^2}{|A_{\rm SM}|^2} + \frac{2\text{Re}(A^*_{\rm SM}A_8)}{|A_{\rm SM}|^2} \Big) + \cdots \Big\}
\end{align}
where $A_6, A_8$ are functions of the Wilson coefficients for dimension-six and eight operators respectively. As we have pulled out an overall factor of the SM amplitude squared, all of the terms in the braces must be dimensionless. We have separated out the scale $\Lambda$ from each term, as this organizes the SMEFT series, so we need to compensate for this scale with powers of the dimensionful scales in the problem -- the Higgs vacuum expectation $v$ or the energy of the process $E$.

\begin{figure}[h!]
    \centering
    \includegraphics[width=\textwidth]{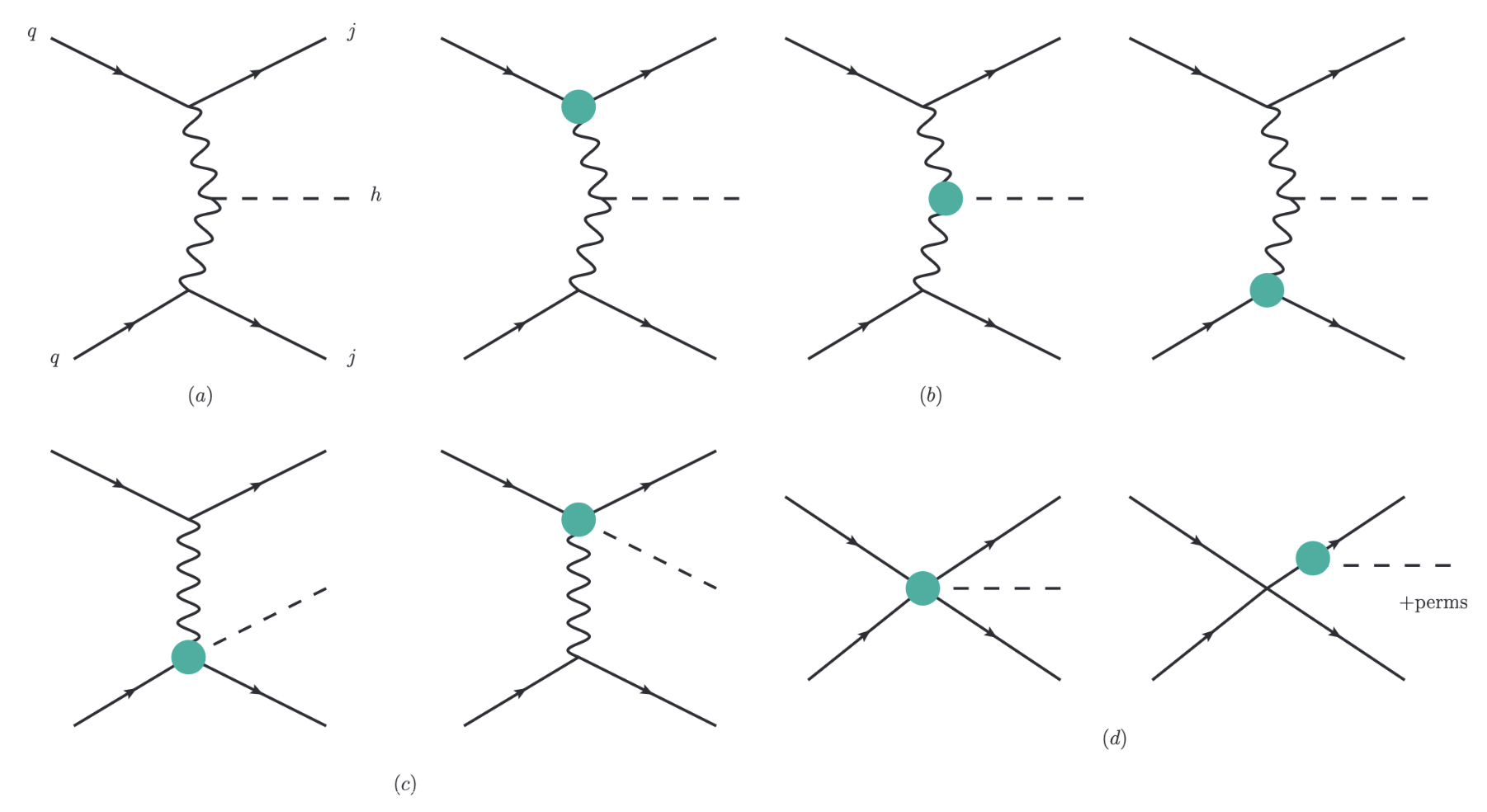}
    \caption{Higgs VBF $t$-channel topologies in the SM and in the SMEFT. The green dots signify modified vertices from the EFT couplings. Corresponding $u$-channel topologies contribute but are implied in the remainder of the text. The figure has been produced with the help of the {\sc JaxoDraw} package~\cite{Binosi:2008ig}}
    \label{fig:vbf-domin-diags}
\end{figure}

Focusing on VBF, let us organize the task of categorizing different energy-enhanced effects by first drawing the possible SM and SMEFT topologies in Fig.~\ref{fig:vbf-domin-diags}. We use the term topology rather than diagram, as several diagrams -- including those with crossed external legs and different types of internal gauge bosons -- are all lumped into the same topology. The top left figure is the SM topology, although SMEFT still affects the SM-type topology by modifying the couplings, i.e., $g_{ffV,\rm SMEFT} = g_{ffV,\rm SM} + \mathcal O(v^2/\Lambda^2)$. This is distinct from the topologies with shaded circles, which represent genuinely new vertex structures. Note that the SMEFT effects from the SM topology are not energy enhanced, always entering with powers of $v^2/\Lambda^2$. A similar set of topologies was shown in Ref.~\cite{Araz:2020zyh}, but our list contains two extra topologies, those labeled as $(c)$ and $(d)$. One of these is the five particle vertices, which cannot be generated at dimension six. As our study includes dimension eight operators such as the class $H^2\psi^4$ shown in left topology in $(d)$. On the other hand we omit the right topology in $(d)$ as small Yukawa coupling factors highly suppress it. Notice further that we have omitted `s-channel' topologies in Fig.~\ref{fig:vbf-domin-diags} where the final state jets come from a vector boson propagator. We have dropped these because the VBF analysis cuts require a large di-jet invariant mass and rapidity gap, which highly suppresses the s-channel contributions~\cite{Araz:2020zyh}.

In the regime where energy \(E\) is significantly greater than the vacuum expectation value \(v\), the terms in \(A_6\) and \(A_8\) that incorporate the highest powers of \(E\) carry the largest impact. These energy-enhanced operators for VBF are provided and discussed in detail later in Tables~\ref{tab:D6_biggest},~\ref{tab:D8_nongeo_ops1},~\ref{tab:D8_nongeo_ops2} and~\ref{tab:D8_nongeo_ops3}. For \(2 \to 3\) particle amplitudes, which inherently have a mass dimension of \(-1\), here is a basic outline of their scaling with energy (while the form for the SM is familiar -- with the factor of $v$ from the $HVV$ vertex -- the scaling for the SMEFT terms will be justified later):
\begin{align}
\label{eq:ecounting}
&\mathcal A_{\rm SM} \sim g^3_{\rm SM}\frac{v}{E^2},\quad
\mathcal A_{Hq}, \mathcal A_{Hu,d} \sim g_{\rm SM} ^2\frac{c_6\, v}{\Lambda^2},
\nonumber\\ 
&\mathcal A_{q^2H^2XD}, \mathcal A_{q^2H^2D^3} \sim g_{\rm SM} ^2\frac{c_8\, v\, E^2}{\Lambda^4},\quad
\mathcal A_{q^4H^2} \sim \frac{c_8\, v\, E^2}{\Lambda^4}
\end{align}
In these expressions, \(g_{\rm SM}\) denotes a SM coupling constant, and \(c_6\) and \(c_8\) are representative Wilson coefficients that arise at each order, indicating the relative importance of different terms as energy scales increase. From these, we can write down the ratio of the dimension-eight interference piece to the dimension-six interference piece
\begin{align}
\label{eq:coeffhierarchy}
\frac{\mathcal{A}^*_{\rm SM}\mathcal{A}_8}{\mathcal{A}^*_{\rm SM}\mathcal{A}_6} \sim \Big(\frac{c_8}{c_6}\Big)\Big(\frac{E^2}{\Lambda^2} \Big).
\end{align}
For a fixed \(\Lambda \sim \text{TeV}\), the Wilson coefficients for energy-enhanced dimension six operators such as \(c^{(3)}_{Hq}\) must be \(\ll 1\) to be consistent with LEP, while dimension-eight contact operators are unconstrained. This allows a hierarchy in couplings, \(c_6 \ll c_8\) (for a given \(\Lambda\)) that upsets the usual power counting, enhancing the effects of dimension-eight by \(c_8/c_6\). Note that we are ignorant of this possibility until we work to \(\mathcal{O}(1/\Lambda^4)\). There are many mechanisms such as in extended gauge group models (e.g. additional $Z'$ bosons or non-Abelian structures)~\cite{Langacker:2008yv}, cancellations among multiple heavy gauge fields reduce dimension-six contributions while leaving dimension-eight operators intact.~\cite{Corbett:2021eux,Dawson:2024ozw}, emphasizing the need for including higher-dimension effects in precision analyses.

Let us plug in some numbers to illustrate this point better. In the kinematic regime where \( E^2/\Lambda^2 \sim 1/5 \), higher-order contributions are typically suppressed by factors of \( E^2/\Lambda^2 \). As a result, the squared amplitudes scale as \( |A|^2_{1/\Lambda^4} \sim 0.2|A|^2_{1/\Lambda^2} \) and \( |A|^2_{1/\Lambda^6} \sim 0.2|A|^2_{1/\Lambda^4} \sim 0.04\,|A|^2_{1/\Lambda^2} \). However, if there is a significant discrepancy between the coefficients \( c_6 \) and \( c_8 \), such as \( c_8/c_6 \sim 5 \), the relation \( |A|^2_{1/\Lambda^2} \sim |A|^2_{1/\Lambda^4} \) can still hold. In this case, the dominant contribution to the overall \( \mathcal{O}(1/\Lambda^4) \) effect comes from the dimension-eight term, as the dimension-six term is relatively suppressed. From this argument, it is clear that the larger the value of \( c_8/c_6 \), the smaller \( E/\Lambda \) needs to be in order for the relation \( |A|^2_{1/\Lambda^2} \sim |A|^2_{1/\Lambda^4} \) to hold. This increases the control over the EFT, as it preserves its predictability for a wider range of energies.

This argument assumes that \( \Lambda \) is not too large, meaning that \( c_6 \sim \mathcal{O}(1) \) would conflict with LEP constraints. However, if we increase \( \Lambda \) to make \( c_6 \sim \mathcal{O}(1) \) consistent with LEP, we end up with a scale beyond the LHC's reach. Therefore, to interpret data from a bottom-up SMEFT perspective with \( \Lambda \) values accessible to the LHC, we need to include dimension-eight operators. Moreover, this assumes that a UV model can generate a small coefficient at dimension six while a larger coefficient at dimension eight. It is well known that, at least for weakly coupled UV scenarios, one can classify operators according to whether they are generated at loop level or tree level  -- though all operators we care about fall into the tree-level category.

Equation~\eqref{eq:coeffhierarchy} is not the end of the story, as the particular kinematics of VBF complicates the energy arguments so that ratios of amplitudes are insufficient, and we must instead compare cross sections. This complication can be traced to forward singularities present in the SM (and some SMEFT) contributions, which cause the cross-section to be dominated by kinematics where $\hat t \ll \hat s$. A hierarchy among kinematic invariants means we need to be more careful about exactly which object enters into Eq.~\eqref{eq:ecounting}. The size and validity range of a given SMEFT operator will clearly depend greatly on its momentum scaling (dictated by the operator's Lorentz, field, and derivative content) and how it compares to the momentum scaling of the SM piece.

As a quick illustration of this point, consider the dimension six operator $Q^2L^2$. If we study how this operator enters the process $q\bar q \to \ell^+\ell^-$, we find
\begin{align}
    |\mathcal A_{\rm SM}|^2 \sim  g^4_{\rm SM}\frac{\hat t^2 + \hat u^2}{\hat s^2},\quad\quad {\rm Re}(A_{\rm SM}A^*_6) \sim\, g^2_{\rm SM}\, c_6\frac{\hat t^2 + \hat u^2}{\hat s\, \Lambda^2},\quad\quad  |A_6|^2 \sim c^2_6 \frac{\hat t^2 + \hat u^2}{\Lambda^4} \nonumber
\end{align}

where $c_6$ is the Wilson coefficient for the operator, $g_{\rm SM}$ is a SM coupling of $O(e)$. For simplicity, we have approximated the SM term with photon exchange and assumed the Lorentz structure of the dimension six operator to be the product of vector currents.
For a fixed $\hat s$, the ratio of amplitudes -- interference term to SM or dimension six squared term to interference term -- is (up to $\mathcal O(1)$ factors) the same as the respective ratio of cross sections. This happens because the scattering angle, the only variable we integrate over two body phase space, does not appear in the denominator. Taken either before or after integration, we find the ratios are $\sim \hat s/\Lambda^2$, as expected from Eq.~\eqref{eq:ecounting}.

Let us contrast this with how the cross sections (for the SM and including $L^2Q^2$) scale for the process $q \ell^- \to q \ell^-$,
\begin{align}
       |\mathcal A_{\rm SM}|^2 \sim  g^4_{\rm SM}\frac{\hat s^2 + \hat u^2}{\hat t^2},\quad\quad {\rm Re}(A_{\rm SM}A^*_6) \sim\, g^2_{\rm SM}\, c_6\frac{\hat s^2 + \hat u^2}{\hat t\, \Lambda^2},\quad\quad  |A_6|^2 \sim c^2_6 \frac{\hat s^2 + \hat u^2}{\Lambda^4}. \nonumber
\end{align}
For fixed $\hat s$, the SM amplitude is dominated by $\hat t \to 0$. Comparing the ratio of amplitudes squared, one would expect the interference $\propto \hat t$ while the squared term is $\propto \hat t^2$. Consequently, for small $\hat t$ , both SMEFT contributions are suppressed. However, the power counting remains consistent: the squared term is suppressed relative to the interference term by the same factor that the interference term is suppressed relative to the SM amplitude squared. If we then integrate the scattering angle $\cos\theta^* = z$ between $[-1+\epsilon, 1-\epsilon]$ and take $\epsilon \ll 1$, we find
\begin{align}
\label{eq:toumodel_tchan}
    \frac{\int dz {\rm Re}(A_{\rm SM}A^*_6)}{\int dz |A_{\rm SM}|^2} \sim \epsilon \log{\epsilon}\frac{\hat s}{\Lambda^2}, \quad\quad  \frac{\int dz |A_6|^2}{\int dz {\rm Re}(A_{\rm SM}A^*_6)} \sim \frac{1}{\log{\epsilon}}\frac{\hat s}{\Lambda^2}.
\end{align}
For small $\epsilon$ -- meaning nearly forward scattering, or large final state $\eta$/low $p_T$ in terms of more collider-friendly observables -- interference term relative the SM square term is suppressed far more than the squared term relative to the interference term by the $epsilon$ dependence.

Extending the example to dimension eight operators of the form $D^2Q^2L^2$, the strength of the various pieces depends on the form of the vertex. If the derivatives are arranged such that the four-fermion vertex $\propto \hat s$, we find the result analogous to Eq.~\eqref{eq:toumodel_tchan} but with $\hat s \to \hat s^2, \Lambda^2 \to \Lambda^4$. However, if the vertex is instead $\propto \hat t$, that factor cancels the $1/\hat t$ in the SM when we square the amplitude. As a result, the ratio of interference (dimension-eight with  SM) relative to SM squared is even more suppressed compared to the ratio of the dimension-eight squared relative to interference
\begin{align}
  \frac{\int dz {\rm Re}(A_{\rm SM}A^*_8)}{\int dz |A_{\rm SM}|^2} \sim \epsilon \frac{\hat s^2}{\Lambda^4}, \quad\quad \frac{\int dz |A_8|^2}{\int dz {\rm Re}(A_{\rm SM}A^*_8)} \sim  \frac{\hat s^2}{\Lambda^4}.
\end{align}

While our example is simplistic in that we picked a specific operator and ignored $\mathcal O(1)$ factors; we still see that i.) we need to compare ratios of cross-section to truly assess the energy enhancement, especially when there is a hierarchy among kinematic invariant, ii.) when SM $\propto 1/\hat t$, interference is suppressed compare to processes where the SM $\propto 1/\hat s$ and depends strongly on the analysis cuts. 

Extrapolating these lessons directly to VBF is not straightforward, as it is a 2 to 3 process, so there are more invariants to consider, but we expect some vestiges of the argument above to persist. Furthermore, in $2 \to 3$ processes (and higher), contributions from SMEFT four-particle vertices are integrated into the full amplitude via propagators. This integration leads to non-trivial momentum dependence in both the numerator — due to energy-enhanced effects —and the denominator. The interplay between the momentum dependencies of the numerator and denominator, along with their interaction with the momentum dependence of the SM amplitude, adds layers of complexity to the calculations.~\footnote{Such numerator and denominator momentum dependencies can occur even in $2 \to 2$ processes,  but they are typically restricted to scenarios involving bosonic initial and final states or dipole operators.}


\section{geoSMEFT and compatible operator bases}\label{sec:geobasis}

In the previous section, we saw that the energy enhancement of a particular SMEFT operator depends on many factors, but it can only arise in operators that have a different operator (Lorentz/derivative) structure than the SM. Therefore, this section aims to identify energy-enhanced operators at $\mathcal O(1/\Lambda^4)$.

The various ways SMEFT operators can enter VBF are shown in Fig.~\ref{fig:vbf-domin-diags}. The effects fall into three different categories: i.) effects of the \(HVV\) vertex, ii.) the \(ffV\) vertex, and contact terms. The \(ffV\) and \(HVV\) effects manifest either as \(v^2/\Lambda^2\) corrections to SM-like vertices or as new Lorentz structures. Only the new Lorentz structures introduce energy enhancements, so we focus on them in this work. The dimension six operators that can enter (either as self-square or interfering among each other) are the same as in Ref.~\cite{Araz:2020zyh} and are listed in Table~\ref{tab:D6_biggest}. We see that $\psi^2H^2D$ affects the $ffV$ vertex (with SM structure) and generates a $ffVH$ contact term.
The other dimension six operators, \(H^2X^2\), impact the \(HVV\) vertex and are indeed energy-enhanced. However, their contributions are sub-leading in VBF because they combine with transverse \(V\) propagators, as discussed in~\cite{Araz:2020zyh}.

\begin{table}[t!]
\center
\begin{tabular}{|r|c|l|}
\hline
&Operator&relevant $\psi$\\
\hline
\hline 
$Q^{(1)}_{H\psi}$&$i(\bar\psi_p\gamma^\nu\psi_r) H^\dag \overleftrightarrow{D}_\mu H$&$\psi=\{q,u,d\}$\\
$Q^{(3)}_{H\psi}$&$i(\bar\psi \gamma^\nu \sigma^I\psi )\, H^\dag \overleftrightarrow{D}_\mu \sigma_I H$&$\psi=\{q\}$\\
\hline
\end{tabular}
\caption{The energy-enhanced dimension six operators which contribute to VBF Higgs production. The remaining $ffV$ and $H VV$ operators were studied for VBF in Ref~\cite{Araz:2020zyh} and explicitly shown to be sub-leading. }\label{tab:D6_biggest}
\end{table}

In order to extend this type of analysis to $\mathcal O(1/\Lambda^4)$, we need to generalize the arguments above on classification and energy-growth for the relevant dimension eight operators involved in VBF Higgs production. The number of operators contributing grows dramatically between dimensions six and eight, so in order to proceed, we require an organizational scheme. 
We start by employing the geometric SMEFT~\cite{Helset:2020yio,Helset:2022pde,Assi:2023zid,Hays:2020scx} to aid in organization and classification. 

In geometric SMEFT, higher dimensional operators are organized according to the particle number of the smallest (in terms of number of particles) vertex they contribute to once all Higgs fields are expanded about their vacuum expectation values (vevs). When dealing with operators with derivatives, the philosophy of the geoSMEFT approach is to strategically place them so that the number of operators that influence two and three-particle vertices is minimized. A few examples of this organizational scheme at work are shown below in Table.~\ref{tab:geosmeft}.
\begin{table}[h!]
\centering
\begin{tabular}{c|c|c|c}

Operator & Minimum Particle Vertex & \# particles & Contribution \\ \hline\hline
$(H^\dag D_\mu H)^*(H^\dag D^\mu H)$ & $v^2 (\partial H)^2$ & 2 & Higgs kinetic term \\ 
$(Q^\dag \bar\sigma_\mu Q)^2$ & $(Q^\dag \bar\sigma_\mu Q)^2$ & 4 & 4-fermion interactions \\ 
$Q\, H\,\sigma_{\mu\nu}\, d\, B^{\mu\nu}$ & $v\,d^\dag_L\,\sigma_{\mu\nu} d_R B^{\mu\nu} + \cdots$ & 3 & Dipole interactions \\ 
$\Box(H^\dag H)\Box(H^\dag H)$ & $(\partial_\mu H^\dag \partial_\mu H \partial^\nu H^\dag \partial_\nu H) + \cdots$ & $4+$ & $4+$ particle vertices \\ & & & after IBP to $\partial_{\mu}\rightarrow D_{\mu}$\\
\end{tabular}
\caption{Examples of geoSMEFT operator organization. In the second column, the $+\cdots$ refers to other four (or more) particle vertices required by gauge invariance.}
\label{tab:geosmeft}
\end{table}
When we restructure the operators, we find that only a small fraction of them contribute to the two- and three-particle vertex category. Specifically, out of 993 dimension-eight operators, only 66 contribute, assuming flavor universality and conservation of baryon and lepton numbers. One important reason the number is small is that these small vertices' kinematics are trivial. By this, we mean that any momentum dot products -- which would come from contracted derivatives in operator language -- can always be expressed as combinations of masses, and therefore, they do not generate any new momentum structure. As a result, the SMEFT expansion for these types of vertex goes from a double expansion in $v/\Lambda$ and $\partial/\Lambda$ to an expansion in $v/\Lambda$ alone.

Furthermore, the number of ways we can add more Higgs fields to an operator, thereby generating more terms in the \(v/\Lambda\) expansion, is restricted by Bose symmetry and the limited number of viable \(SU(2)\) tensor structures. Consequently, it is possible to determine the number and form of operators contributing to 2- and 3-particle vertices to all orders in \(v/\Lambda\). In Ref.~\cite{Helset:2020yio}, these vertices were referred to as Higgs-dependent ``metrics" and were derived  to all orders. For example, denoting the four real degrees of freedom in the Higgs doublet as \(\phi^I\), \(I = 1\cdots 4\), the Higgs kinetic term metric is \(h_{IJ}(\phi)D_\mu \phi^I D^\mu \phi^J\), where
\begin{align}
h_{IJ}={}&\left[1+\phi^2c_{H\Box}^{(6)}+\sum_{n=0}^\infty\big(\frac{\phi^2}{2}\big)^{n+2}\big(c_{HD}^{(8+2n)}-c_{HD,2}^{(8+2n)}\big)\right]\delta_{IJ}\nonumber\\
{}&+\frac{\Gamma_{A,J}^I\phi_K\Gamma^K_{A,L}\phi^L}{2}\big(\frac{c_{HD}^{(6)}}{2}+\sum_{n=0}^\infty\big(\frac{\phi^2}{2}\big)^{n+1}c_{HD,2}^{(8+2n)}\big)\, .\label{eq:hmetric}
\end{align}
The $\Gamma^I_{AJ}$ represent different formulations of the $SU(2)_L \times U(1)_Y$ generators adapted to the four-component fields $\phi$ and $W$ (combining the $SU(2)$ and hypercharge gauge fileds together), and $c^{(6)}_{H\Box}, c^{(6)}_{HD}, c^{8+2n}_{HD},c^{8+2n}_{HD,2}$ are the Wilson coefficients of the operators:
\begin{align}
& Q^{(6)}_{H\Box} = (H^\dag H)\Box(H^\dag H), \nonumber \\ 
& Q^{(6)}_{HD} = (H^\dag D^\mu H)^*(H^\dag\,D_\mu H), \nonumber \\ 
& Q_{H D}^{(8+2 n)}=\big(H^{\dagger} H\big)^{n+2}\big(D_\mu H\big)^{\dagger}\big(D^\mu H\big), \nonumber \\
& Q_{HD, 2}^{(8+2 n)}=\big(H^{\dagger} H\big)^{n+1}\big(H^{\dagger} \sigma_a H\big)\big(D_\mu H\big)^{\dagger} \sigma^a\big(D^\mu H\big). \nonumber
\end{align}
Notice that only two new operators enter at each mass dimension $8+2n$. Analogous expressions for all other 2-particle and 3-particle metrics were shown in Ref.~\cite{Helset:2020yio}.

For the purposes of this work, the most essential aspect of the geoSMEFT approach is how it facilitates energy counting. As mentioned earlier, the metrics such as $h_{IJ}$ are functions of $\phi \sim \sqrt{H^\dag H}$ alone, i.e., lacking covariant derivatives, so higher order terms in the metrics lead to an expansion purely in $v^2/\Lambda^2$ only, rather than mixed powers of $v$ and energy. More specifically, the powers of energy a particular three-point vertex carries is entirely set by the lowest dimension operator that generates the vertex. With our assumptions of baryon/lepton number conservation, the lowest dimension operator (for a two- or three-particle vertex) is either dimension four (for SM-like kinematics) or dimension six (new kinematics). At dimension eight and higher, SMEFT operators will only introduce additional factors of $v^2$. All energy-enhanced effects are shuffled into operators that only affect four (or higher) particle vertices.  

While the number of operators that affect four-particle vertices is large at any given dimension, they are process-dependent. For any given process, say \(q\bar q \to h jj\), the number of operators that can contribute (at tree level) is much smaller. Additionally, these operators need to interfere with the SM, meaning the dimension-eight operators must have the same chirality, color structure, and Lorentz structure. This requirement shrinks the number of operators significantly. 

Given the arguments above, what we want to find are the operators contributing to four or more particle vertices with the proper structure to interfere with the SM. The geoSMEFT organizes all operators impacting two- and three-point functions in the SMEFT and partially covers operators affecting four- and higher-point functions. These include, for example, \(\psi^2H^2D\), which contains the \(qqVH\) vertex, but to find the complete set of operators, we have to look beyond the geoSMEFT operator set. The procedure, first implemented in Ref.~\cite{Corbett:2023yhk} for operators relevant for VBF and $VH$ production, involves revising the dimension-eight operator basis by identifying all operators affecting two- and three-point functions not included in the geoSMEFT, removing these through integration by parts and equations of motion and incorporating the resulting new operator forms~\cite{Kim:2022amu}. For example, the operator~\cite{Li:2020gnx}, 
\begin{align}
Q^{(1)}_{H^2H^{\dagger 2}D^4} = (H^\dag H)\Box^2(H^\dag H)
\end{align}
which becomes relevant post-spontaneous symmetry breaking, is inconsistent with the geoSMEFT basis but can be substituted with operators from Ref.~\cite{Murphy:2020rsh} that align with the geoSMEFT and do not generate two- or three-point functions due to their complete reliance on derivatives~\footnote{Explicitly, $Q^{(1)}_{H^4} = (D_\mu H^\dag D_\nu H)(D^\nu H^\dag D^\mu H)$, $Q^{(2)}_{H^4} = (D_\mu H^\dag D_\nu H)(D^\mu H^\dag D^\nu H)$ and $Q^{(3)}_{H^4} = (D^\mu H^\dag D_\mu H)(D^\nu H^\dag D^\nu H)$.}. All operators in this study are presented in the geoSMEFT basis as described in Ref. ~\cite{Corbett:2023yhk}. 

\begin{table}[t!]
\center
\begin{tabular}{|r|c|l|}
\hline
&Operator&relevant $\psi$\\
\hline
\hline
$Q_{\psi^2BH^2D}^{(1)}$&$(\bar\psi_p\gamma^\nu\psi_r)D^\mu(H^\dagger H)B_{\mu\nu}$&$\psi=\{q,u,d\}$\\
$Q_{\psi^2BH^2D}^{(2)}$&$i(\bar\psi_p\gamma^\nu\psi_r)(H^\dagger \overleftrightarrow D^\mu H)B_{\mu\nu}$&$\psi=\{q,u,d\}$\\
$Q_{\psi^2BH^2D}^{(3)}$&$(\bar\psi_p\gamma^\nu\sigma^I\psi_r)D^\mu(H^\dagger \sigma^IH)B_{\mu\nu}$&$\psi=\{q\}$\\
$Q_{\psi^2BH^2D}^{(4)}$&$i(\bar\psi_p\gamma^\nu\sigma^I\psi_r)(H^\dagger \overleftrightarrow D^{I\mu}H)B_{\mu\nu}$&$\psi=\{q\}$\\
\hline
$Q_{\psi^2WH^2D}^{(1)}$&$(\bar\psi_p\gamma^\nu\psi_r)D^{\mu}(H^\dagger \sigma^IH)W_{\mu\nu}^I$&$\psi=\{q,u,d\}$\\
$Q_{\psi^2WH^2D}^{(2)}$&$i(\bar\psi_p\gamma^\nu\psi_r)(H^\dagger \overleftrightarrow D^{I\mu}H)W_{\mu\nu}^I$&$\psi=\{q,u,d\}$\\
$Q_{\psi^2WH^2D}^{(3)}$&$(\bar\psi_p\gamma^\nu\sigma^I\psi_r)D^\mu(H^\dagger H)W_{\mu\nu}^I$&$\psi=\{q\}$\\
$Q_{\psi^2WH^2D}^{(4)}$&$i(\bar\psi_p\gamma^\nu\sigma^I\psi_r)(H^\dagger \overleftrightarrow D^\mu H)W_{\mu\nu}^I$&$\psi=\{q\}$\\
$Q_{\psi^2WH^2D}^{(5)}$&$\epsilon_{IJK}(\bar\psi_p\gamma^\nu\sigma^I\psi_r)D^\mu(H^\dagger \sigma^JH)W_{\mu\nu}^K$&$\psi=\{q\}$\\
$Q_{\psi^2WH^2D}^{(6)}$&$i\epsilon_{IJK}(\bar\psi_p\gamma^\nu\sigma^I\psi_r)(H^\dagger \overleftrightarrow D^{J\mu}H)W_{\mu\nu}^K$&$\psi=\{q\}$\\
\hline
\end{tabular}
\caption{Operators of class $\psi^2XH^2D$ contributing to Higgs associated production up to $\mathcal O(1/\Lambda^4)$. The last column indicates which chiral quarks contribute. Note our naming convention differs from that of \cite{Murphy:2020rsh} and our convention for $\overset\leftrightarrow{D}$ differs by a factor of $i$.}\label{tab:D8_nongeo_ops1}
\end{table}

\begin{table}[t!]
\center
\begin{tabular}{|r|c|l|}
\hline
&Operator&relevant $\psi$\\
\hline
\hline
$Q_{\psi^2H^2D^3}^{(1)}$&$i(\bar\psi_p\gamma^\mu\psi_r)\left[(D_\nu H)^\dagger (D^2_{(\mu,\nu)}H)-(D^2_{(\mu,\nu)}H)^\dagger (D_\nu H)\right]$&$\psi=\{q,u,d\}$\\
$Q_{\psi^2H^2D^3}^{(2)}$&$i(\bar\psi_p\gamma^\mu \overleftrightarrow D_\nu\psi_r)\left[(D_\mu H)^\dagger(D_\nu H)+(D_\nu H)^\dagger(D_\mu H)\right]$&$\psi=\{q,u,d\}$\\
$Q_{\psi^2H^2D^3}^{(3)}$&$i(\bar\psi_p\gamma^\mu\sigma^I\psi_r)\left[(D_\nu H)^\dagger \tau^I(D^2_{(\mu,\nu)}H)-(D^2_{(\mu,\nu)}H)^\dagger\sigma^I (D_\nu H)\right]$&$\psi=\{q\}$\\
$Q_{\psi^2H^2D^3}^{(4)}$&$i(\bar\psi_p\gamma^\mu \sigma^I\overleftrightarrow D_\nu\psi_r)\left[(D_\mu H)^\dagger\tau^I(D_\nu H)+(D_\nu H)^\dagger\tau^I(D_\mu H)\right]$&$\psi=\{q\}$\\
\hline
\end{tabular}
\caption{Operators of class $\psi^2H^2D^3$ contributing to Higgs associated production up to $\mathcal O(1/\Lambda^4)$. The last column indicates which chiral quarks contribute. The subscript $(\mu,\nu)$ indicates that the derivatives are symmetrized in the Lorentz indices.}\label{tab:D8_nongeo_ops2}
\end{table}

\begin{table}[t!]
\center
\begin{tabular}{|r|c|}
\hline
&Operator\\
\hline
\hline
$Q_{q^4H^2}^{(1)}$&$ (\bar q_p\gamma^\mu q_r)(\bar q_p\gamma_\mu q_r)(H^\dag H) $ \\
$Q_{q^4H^2}^{(2)}$&$  (\bar q_p\gamma^\mu q_r)(\bar q_p\gamma_\mu \sigma^I q_r)(H^\dag \sigma^I H)$ \\
$Q_{q^4H^2}^{(3)}$&$ (\bar q_p\gamma^\mu \sigma^I q_r)(\bar q_p\gamma_\mu \sigma^I q_r)(H^\dag H) $\\
$Q_{u^4H^2}^{(1)}$&$ (\bar u_p\gamma^\mu u_r)(\bar u_p\gamma_\mu u_r)(H^\dag H) $ \\
$Q_{d^4H^2}^{(1)}$&$ (\bar d_p\gamma^\mu d_r)(\bar d_p\gamma_\mu d_r)(H^\dag H) $ \\
$Q_{u^2d^2H^2}^{(1)}$&$ (\bar u_p\gamma^\mu u_r)(\bar d_p\gamma_\mu d_r)(H^\dag H) $ \\
$Q_{q^2u^2H^2}^{(1)}$&$ (\bar q_p\gamma^\mu q_r)(\bar u_p\gamma_\mu u_r)(H^\dag H) $ \\
$Q_{q^2u^2H^2}^{(2)}$&$ (\bar q_p\gamma^\mu \sigma^I q_r)(\bar u_p\gamma_\mu u_r)(H^\dag\sigma^I H) $ \\
$Q_{q^2d^2H^2}^{(1)}$&$ (\bar q_p\gamma^\mu q_r)(\bar d_p\gamma_\mu d_r)(H^\dag H) $ \\
$Q_{q^2d^2H^2}^{(2)}$&$ (\bar q_p\gamma^\mu \sigma^I q_r)(\bar d_p\gamma_\mu d_r)(H^\dag\sigma^I H) $ \\
\hline
\end{tabular}
\caption[Caption for LOF]{Operators of class $\psi^4H^2$ contributing to Higgs associated production up to $\mathcal O(1/\Lambda^4)$. There are variations of many of these operators containing different color contractions (products of octet currents), these operators do not interfere with the (color-singlet) SM VBF structure\protect\footnotemark. Similarly, four fermion operators of the form $(\bar L R)(\bar L R) + h.c.$ or $(\bar L R)(\bar R L)$ -- where $L/R$ refer to the handedness of the fermions involved and spinor indices are contracted within the parenthesis -- have the wrong helicity structure to interfere with the SM.}
\label{tab:D8_nongeo_ops3}
\end{table}
\footnotetext{This assertion is operator basis-dependent. If one chooses a  different color structure rather than forming octet currents, these operators indeed interfere with the SM amplitude.}

The operators listed in Tables~\ref{tab:D8_nongeo_ops1} and~\ref{tab:D8_nongeo_ops2} were already examined in Ref.~\cite{Corbett:2023yhk} in the context of $VH$ production. This overlap is expected because the processes are related by crossing symmetry. In that reference, it was shown that these tables include all the dimension‑eight operators that can interfere with the SM in $VH$ production. At first glance, one might think that imposing flavor and CP assumptions is unnecessary since other operator classes are naturally excluded due to a mismatch with the helicity structure of the SM. However, flavor and CP assumptions are crucial to address two types of dimension‑6 squared contributions, which are formally of the same SMEFT order as the dimension‑8 SM interference terms. 

The first type, arising from the square of the amplitude with one dimension‑6 vertex insertion, does not interfere with the SM amplitude. Such contributions, which do not align with the SM structure, are suppressed by the assumed hierarchy \( c^{(6)} \ll c^{(8)} \). The second type involves the interference between the SM amplitude and the amplitude with two dimension‑6 vertex insertions. This interference is naturally suppressed since two insertions of \( c^{(6)} \) lead to a smaller effect than a single insertion of \( c^{(8)} \), and the assumption of minimal flavor violation further constrains the flavor structure of the Wilson coefficients. To control the proliferation of dimension‑6 squared terms and maintain a manageable analysis, we impose CP conservation and \( U(3)^5 \) flavor symmetry in Tables~\ref{tab:D8_nongeo_ops1} and~\ref{tab:D8_nongeo_ops2}. These assumptions ensure that the analysis focuses solely on the relevant dimension‑8 operators without being overwhelmed by contributions from dimension‑6 squared terms.

The operators in Table~\ref{tab:D8_nongeo_ops3}, illustrated in Fig.~\ref{fig:vbf-domin-diags} (d) can modify VBF by new contact interactions between quarks and the Higgs field, altering observables like the transverse momentum of the jets and the Higgs, as well as affecting the total cross section and angular distributions in VBF processes.
These operators did not appear in Ref.~\cite{Corbett:2023yhk} because the authors assumed the vector boson in VH production decayed leptonically.
While semi-leptonic versions of Table~\ref{tab:D8_nongeo_ops3}, operators in class, e.g. $L^2q^2H^2$, could play a role, Ref.~\cite{Corbett:2023yhk} assumed their effect would be suppressed by cuts on the dilepton invariant mass aimed to select out resonant $V \to \ell \ell$ production. We will explore the impact of the operators of Table~\ref{tab:D8_nongeo_ops3} in $VH$ more thoroughly in Sec.~\ref{sec:associated}.

\section{Energy-enhanced contributions to VBF}\label{sec:SMEFTcont}

In this section, we study $qV \to q'H$, a two-to-two process we can use as a simple proxy for VBF to better understand which operators introduced in the last section are most important. This is the same calculation from Ref.~\cite{Araz:2020zyh}, extended to $\mathcal O(1/\Lambda^4)$ (including both dimension-six squared and dimension-eight operators). While this expanded calculation supports the energy counting arguments in Section~\ref{sec:dim8ops}, we further find that the large $t$ suppression argument in Ref.~\cite{Araz:2020zyh} is not entirely valid and depends more intricately on the operator structures involved. 

We begin by considering the $2\to 2$ amplitude for left-handed (LH) quarks, including the SM, as illustrated below in Fig.~\ref{fig:vbf-22}.
\begin{figure}[tb]
    \centering
    \includegraphics[width=0.9\textwidth]{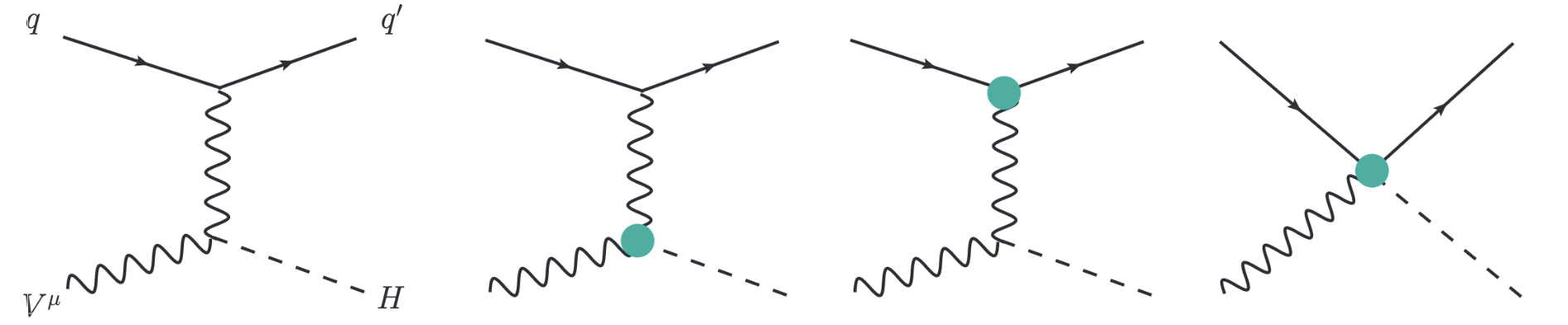}
    \caption{Higgs VBF $2\to 2$ sub-topologies. The figure has been produced with the help of the {\sc JaxoDraw} package~\cite{Binosi:2008ig}}
    \label{fig:vbf-22}
\end{figure}
We take $V=Z$, for starters, taking all momenta outgoing:
\begin{align}
{\cal A}(qZ_{L,\mu}\rightarrow q H) &= i\bra{\bar q}\gamma^{\mu}|q]\Big( g_{Zq_Lq_L}\,g^{(1)}_{HZZ}\, \hat p_{q\bar q,Z} + g_{Zq_Lq_L}\,g^{(2)}_{HZZ}\, \hat p_{q\bar q,Z}\, (p_H \cdot p_V + m^2_V) + g^{(1)}_{ZHq_Lq_L} + \nonumber \\
& g^{(2)}_{ZHq_Lq_L}(p_H\, \cdot p_Z)\Big) + i\bra{\bar q}\slashed p_H |q]\, p^\mu_H\, \Big( g_{Zq_Lq_L}\,g^{(2)}_{HZZ}\, \hat p_{q \bar q,Z} + g^{(3)}_{ZHq_Lq_L} \Big)
\label{eq:2to2LH}
\end{align}
where $p_{23,V}$ contains a propagator, and the translation between couplings and SMEFT Wilson coefficients are given in Table~\ref{tab:couplingsZ},
\begin{table}[h!]
\centering
\footnotesize
\renewcommand{\arraystretch}{1.4}
\begin{tabular}{ll}
\hline
\textbf{Coupling} & \textbf{Expression} \\
\hline
\multicolumn{2}{c}{\textbf{Dimension-Six}} \\
\hline
\(g_{HZZ}^{(2)}\) & \(4v\left[ c_w^2\, c_{HW} + s_w\, c_w\, c_{HWB} + s_w^2\, c_{HB} \right]\) \\[1.5ex]
\(g_{ZHu_Lu_L}^{(1)}\) & \(\displaystyle \frac{e v}{c_w s_w}\big( c_{Hq}^{(3)} - c_{Hq}^{(1)} \big)\) \\[2.5ex]
\(g_{ZHd_Ld_L}^{(1)}\) & \(\displaystyle -\frac{e v}{c_w s_w}\big( c_{Hq}^{(3)} + c_{Hq}^{(1)} \big)\) \\[2.5ex]
\hline
\multicolumn{2}{c}{\textbf{Dimension-Eight}} \\
\hline
\(g_{ZHu_Lu_L}^{(2)}\) & \(\displaystyle \frac{v}{4 c_w s_w} \left[ -e \left( c^{(1)}_{q^2H^2D^3} - c^{(3)}_{q^2H^2D^3} \right) \right. \) \\[1ex]
& \(\displaystyle \left. +\, 4 c_w s_w \left( s_w \left( c^{(3)}_{q^2BH^2D} - c^{(1)}_{q^2BH^2D} \right) + c_w \left( c^{(1)}_{q^2WH^2D} - c^{(3)}_{q^2WH^2D} \right) \right) \right] \) \\[2.5ex]
\(g_{ZHd_Ld_L}^{(2)}\) & \(\displaystyle \frac{v}{4 c_w s_w} \left[ 4 c_w s_w \left( s_w \left( c^{(1)}_{q^2BH^2D} + c^{(3)}_{q^2BH^2D} \right) \right. \right. \) \\[1ex]
& \(\displaystyle \left. \left. +\, c_w \left( c^{(1)}_{q^2WH^2D} + c^{(3)}_{q^2WH^2D} \right) \right) - e \left( c^{(1)}_{q^2H^2D^3} + c^{(3)}_{q^2H^2D^3} \right) \right] \) \\[2.5ex]
\(g_{ZHu_Lu_L}^{(3)}\) & \(\displaystyle \frac{v}{4 c_w s_w} \left[ 3 e \left( c^{(1)}_{q^2H^2D^3} - c^{(3)}_{q^2H^2D^3} \right) \right. \) \\[1ex]
& \(\displaystyle \left. +\, 4 c_w s_w \left( s_w \left( c^{(3)}_{q^2BH^2D} - c^{(1)}_{q^2BH^2D} \right) + c_w \left( c^{(1)}_{q^2WH^2D} - c^{(3)}_{q^2WH^2D} \right) \right) \right] \) \\[2.5ex]
\(g_{ZHd_Ld_L}^{(3)}\) & \(\displaystyle \frac{v}{4 c_w s_w} \left[ 3 e \left( c^{(1)}_{q^2H^2D^3} + c^{(3)}_{q^2H^2D^3} \right) \right. \) \\[1ex]
& \(\displaystyle \left. +\, 4 c_w s_w \left( s_w \left( c^{(1)}_{q^2BH^2D} + c^{(3)}_{q^2BH^2D} \right) + c_w \left( c^{(1)}_{q^2WH^2D} + c^{(3)}_{q^2WH^2D} \right) \right) \right] \) \\[2.5ex]
\hline
\end{tabular}
\caption{Couplings separated into dimension-six and dimension-eight contributions, with implicit dependence on \(\Lambda\).}
\label{tab:couplingsZ}
\end{table}

We further use the shorthand
\begin{equation}
    \hat{p}_{ij,V}=\frac{1}{(p_i+p_j)^2-m_V^2+i\Gamma_Vm_V}
\end{equation}
and kinematic variables $s_{ij}=(p_i+p_j)^2$ throughout this section. Within Eq.~\eqref{eq:2to2LH}, the coupling factors $g_{Zqq}$ and $g^{(1)}_{HZZ}$ begin (in terms of minimum powers of $\Lambda$) at $\mathcal O(\Lambda^0)$, $g^{(2)}_{HZZ}$ and the four-particle contact term $g^{(1)}_{ZHqq}$ begin at $\mathcal O(\Lambda^{-2})$, while the remaining four particle contact terms $g^{(2)}_{ZHqq}, g^{(3)}_{ZHqq}$ begin at $\mathcal O(\Lambda^{-4})$. For coupling factors that have a minimum order less than $\mathcal O(\Lambda^{-4})$, we will only keep the leading (fewest powers of $\Lambda$) piece. For example, while there are $\mathcal O(\Lambda^{-2})$ and $\mathcal O(\Lambda^{-4})$ contributions to the $ffV$ couplings $g_{Zqq}$, we will neglect them. We do this because we focus on the effects with the largest energy enhancement, and higher order (in $\Lambda$) corrections are always of the form $(v/\Lambda)^n$.  

The analog of Eq.~\eqref{eq:2to2LH} with right-handed (RH) quarks can be derived by replacing LH couplings with RH versions as shown in Table~\ref{tab:rem_couplings_all} and swapping $\bra{\bar q}\gamma^{\mu}|q]\rightarrow [\bar{q}|\gamma^{\mu}\ket{q}$. A similar replacement gets us $\bar q V \to \bar q H$. We focus on the quark-initiated variation here because $q q' \to q q' H$ is significantly larger than the other VBF subprocesses. Within $qq' \to qq'H$, the helicity combination $q_L q'_L \to q_L q'_L H$ is the largest, as it proceeds via both $W$ fusion and $Z$ fusion. Moving beyond the SM, there could, in principle, be contributions from vertices involving one left-handed fermion and one right-handed fermion (LR or RL) (rather than two) left (LL) or two right (RR) chirality fermions as in the SM, such as arise in SMEFT from dipole operators. However, we will ignore this possibility here as our assumption of flavor universality removes all (LR/RL) structure operators.

To better illustrate which effects grow most strongly with energy, we consider the limit where \(2(p_H \cdot p_V) \approx 2(p_q \cdot p_{\bar{q}}) \approx \hat{t}\gg m_V\). We also assume that the incoming vector boson \(V\) is longitudinally polarized. This assumption is justified because, in the Standard Model (SM), the dominant contribution to the vector boson fusion (VBF) process arises from longitudinally polarized vector bosons once the subprocess \(q V \to \bar{q} H\) is embedded into the full VBF process~\cite{Dawson:1984gx}. Mathematically, this amounts to contracting the polarization vector \(\epsilon^\mu(p_V) \propto p^\mu_V\) into Eq.~\eqref{eq:2to2LH} and neglecting the masses of the vector boson \(V\) and the Higgs boson \(H\). By performing this contraction and taking the limit \(\hat{t} \gg m_V^2, m_H^2\), reinstating the minimum powers of \(\Lambda\), we obtain,
\begin{align}
{\cal A}(qZ_{L,\mu}\rightarrow q H) \underset{{\hat t \gg m^2_{Z,H}}}{=} -i\bra{\bar q}\slashed p_H |q] \frac{1}{\hat t}\Big (g_{Zq_Lq_L}g^{(1)}_{HZZ} + g^{(1)}_{ZHq_Lq_L} \frac{\hat t }{\Lambda^2} + (g^{(2)}_{ZHq_Lq_L} - g^{(3)}_{ZHq_Lq_L}) \frac{\hat t^2 }{2\Lambda^4}\Big).
\label{eq:2to2largeT}
\end{align}

In this expression, the first term involving \(g_{Zq_Lq_L}g^{(1)}_{HZZ}\) corresponds to the SM contribution. The subsequent terms arise from higher-dimensional operators in the SMEFT framework. Importantly, we observe that the terms most enhanced in the large \(\hat{t}\) limit are the four-particle contact terms, which scale with higher powers of \(\hat{t}\). Up to \(\mathcal{O}(\Lambda^{-2})\), this result reproduces the calculation presented in Ref.~\cite{Araz:2020zyh}. However, we find that the large \(\hat{t}\) suppression argument from the reference is not entirely valid. The suppression depends more intricately on the specific operator structures involved as we will see later.

One may expect the \(\mathcal{O}(\Lambda^{-4})\) correction is always small, at least in the limit where the EFT is valid. However, as alluded to in Sec.~\ref{sec:dim8ops}, coefficient hierarchies -- here in the form $g^{(1)}_{ZHqq} \ll g^{(2,3)}_{ZHqq} $ can upset this line of thinking. Note that $g^{(1)}_{HZZ}, g^{(i)}_{ZHqq}$ are all $\propto v$, which is needed to keep the mass dimensions in Eq.~\eqref{eq:2to2largeT} correct. Note that operators in the class $\psi^2XH^2D$ operators drop of Eq.~\eqref{eq:2to2largeT} as they do not contain to longitudinal $V$ in the high energy limit.

From Eq.~\eqref{eq:2to2LH}, we can justify the energy scalings we presented in Eq.~\eqref{eq:coeffhierarchy}. As shown in Table~\ref{tab:couplingsZ}, all the couplings are proportional to \( v \). For massless fermions, the spinors scale as \( | i \rangle, | i ] \sim \sqrt{E} \), so each fermion current contributes a factor proportional to \( E \). Each propagator scales as \( 1/E^2 \). Combining these factors for the \( 2 \to 2 \) sub-amplitude \( qV \to q' H \), we find:
\begin{align}
    A_{\text{SM}} \sim \frac{v}{E},\quad A_{\text{dim-6}} \sim \frac{v\, E}{\Lambda^2},\quad A_{\text{dim-8}} \sim \frac{v\, E^3}{\Lambda^4}
\end{align}
To extend this scaling to the full VBF amplitude, we include an additional factor of \( 1/E \). This arises from the quark current (providing a factor of \( E \)) and the vector boson propagator (providing \( 1/E^2 \)) that connect the subprocess to the rest of the diagram. Including this \( 1/E \) factor and the coupling constants, we recover the scaling behavior presented in Eq.~\eqref{eq:coeffhierarchy}.

Next, let us repeat the exercise above, but with an initial state $W$-boson instead of a $Z$. The amplitude for the process $q W_\mu \rightarrow q' H$ is given by: 
\begin{align}
{\cal A}(qW_\mu\rightarrow q' H) &= i\bra{\bar q}\gamma^{\mu}|q]\Big( g_{Wq_Lq'_L}\,g^{(1)}_{HWW}\, \hat p_{qq',W} + g_{Wq_Lq'_L}\,g^{(2)}_{HWW}\, \hat p_{qq',W}\, (p_H \cdot p_W + m^2_W)  + \nonumber \\
& g^{(1)}_{WHq_Lq'_L}+ g^{(2)}_{WHq_Lq'_L}(p_H\, \cdot p_W)  + g^{(3)}_{WHq_Lq'_L}(p_H\, \cdot (p_q - p_{q'}) \Big) + \nonumber \\
& i\bra{\bar q}\slashed p_H |q]\, \Big( p^\mu_H ( g_{Wq_Lq'_L}\,g^{(2)}_{HWW}\, \hat p_{q q',W} + g^{(4)}_{WHq_Lq'_L}) + g^{(3)}_{WHq_Lq'_L}\,(p^\mu_q - p^\mu_{q'}) \Big)
\label{eq:2to2LHa}
\end{align}
Note that, in comparison with Eq.~\eqref{eq:2to2LH}, additional terms appear that involve the momenta of the quarks contributing to the amplitude. Continuing as in the $Z$ exchange case and inserting the polarization vector for a longitudinal $W$, we find.
\begin{align}
{\cal A}(qW_{L,\mu}\rightarrow {q}' H) \underset{{\hat t \gg m^2_{W,H}}}{=}& -i\bra{\bar q}\slashed p_H |q] \frac{1}{\hat t}\Big (g_{Wq_Lq'_L}g^{(1)}_{HWW} + g^{(1)}_{WHq_Lq'_L} \frac{\hat t }{\Lambda^2} + (g^{(2)}_{WHq_Lq'_L} - g^{(4)}_{WHq_Lq'_L} ) \frac{\hat t^2 }{2\Lambda^4} \nonumber\\
&- g^{(3)}_{WHq_Lq'_L}  \frac{\hat t\,(2\hat s + \hat t) }{2\Lambda^4}  \Big).
\label{eq:2to2largeTa}
\end{align}
with couplings given in Table~\ref{tab:W_couplings}.
\begin{table}[tb]
\footnotesize
\centering
\renewcommand{\arraystretch}{1.6}
\begin{tabular}{ll}
\hline
\textbf{Coupling} & \textbf{Expression} \\
\hline
\multicolumn{2}{c}{\textbf{Dimension-Six}} \\
\hline
\(g_{WH\bar{u}_L d_L}^{(1)}\) & \(\displaystyle \sqrt{2}\,\frac{e v}{s_w}\, c_{Hq}^{(3)}\, V_{ud}\) \\[1.5ex]
\(g_{HWW}^{(2)}\) & \(\displaystyle 4 v\, c_{HW}\) \\[1.5ex]
\hline
\multicolumn{2}{c}{\textbf{Dimension-Eight}} \\
\hline
\(g_{WH\bar{u}_L d_L}^{(2)}\) & \(\displaystyle \frac{v}{2 \sqrt{2}\, s_w}\, V_{ud} \left[  e\, c^{(3)}_{q^2 H^2 D^3} - 4 s_w \big( c^{(3)}_{q^2 W H^2 D} + i\,c^{(3)}_{q^2 W H^2 D} \big) \right] \) \\[1.5ex]
\(g_{WH\bar{u}_L d_L}^{(3)}\) & \(\displaystyle \frac{e v}{\sqrt{2}\, s_w}\, V_{ud}\, c^{(4)}_{q^2 H^2 D^3} \) \\[1.5ex]
\(g_{WH\bar{u}_L d_L}^{(4)}\) & \(\displaystyle - \frac{v}{2 \sqrt{2}\, s_w}\, V_{ud} \left[ 3 e\, c^{(3)}_{q^2 H^2 D^3} + 4 s_w \big( c^{(3)}_{q^2 W H^2 D} + i c^{(5)}_{q^2 W H^2 D} \big) \right] \)\\ [1.5ex]
\hline
\end{tabular}
\caption{Couplings separated into dimension-six and dimension-eight contributions, with implicit dependence on \(\Lambda\).}
\label{tab:W_couplings}
\end{table}
A key observation is the presence of terms proportional to $\hat{s}\hat{t}$ relative to the SM. In the regime where $\hat{s}$ is large but $\hat{t}$ remains small, these $\hat{s}\hat{t}$ terms dominate while other SMEFT contributions are suppressed by $\hat{t}$~\footnote{When $\hat{t}$ is small, one should also retain terms of $\mathcal{O}(m_W^2)$}. Squaring Eq.\eqref{eq:2to2LH} (or Eq.~\eqref{eq:2to2largeTa}) shows that SMEFT terms proportional to $\hat{t}$ result in differential cross sections with positive powers of $\hat{t}$, as the SM’s $1/\hat{t}$ is canceled by the $\hat{t}$ enhancement. In contrast, SMEFT terms proportional to $\hat{s}$ or $\hat{s}^2$ lead to enhancements scaling with $\hat{s}$ or $\hat{s}^2$.

To further quantify how these kinematic differences appear in a complete two-to-three process, we utilize the effective $W$ approximation~\cite{Dawson:1984gx}, treating the incoming $W$ boson as a constituent of the proton. In this framework, the $W$ distribution function is convolved with the two-to-two cross-section $\hat{\sigma}(q W \to q H)$, calculated in the small-angle limit (i.e., $\hat{t} \to 0$). The $W$ distribution functions depend on the polarization of the $W$ boson, with the transverse parton distribution function (pdf) being significantly larger than the longitudinal one.

Integrating over the scattering angle up to the maximum value of \(\theta^*_{\text{max}} = p_{T, W}/E_W \sim 2 m_W/\sqrt{\hat{s}}\), and taking the limit of large \(\hat{s}\), the SM \(q W \to q' H\) amplitude squared for longitudinal \(W\) is larger than for transverse \(W\) by a factor of \(\mathcal{O}(\hat{s}/m_W^2)\). This factor offsets the smaller effective parton distribution function for longitudinal \(W\), resulting in the SM amplitude being dominated by the longitudinal \(W\) component.

For the dimension-six term \(g^{(1)}_{WHq_Lq_L}\) (which corresponds to the operator \(c_{Hq}^{(3)}\) from Table~\ref{tab:W_couplings}), both the interference and self-squared terms are inversely proportional to \(\hat{s}\) after the \(\theta^*\) integration when the incoming \(W\) is transverse. They are, therefore, suppressed in the large \(\hat{s}\) limit. For the longitudinal case, we find
\begin{align}
\label{eq:effW_dim6}
    \int^{\theta_{\text{max}}} d\theta^*\, 2\,\text{Re}(A_{\text{SM}} A^{(6)})_{W_L} \sim \frac{v^2\, \hat{s}}{\Lambda^2\, m_W^2},\quad
    \int^{\theta_{\text{max}}} d\theta^*\, |A^{(6)}|^2_{W_L} \sim \frac{v^2\, \hat{s}}{\Lambda^4},
\end{align}
where we have focused on the scaling with dimensionful quantities and omitted all coupling constants and \(\mathcal{O}(1)\) factors.

Repeating the calculation for dimension-eight operators, the $\hat s\,$ vs.  $\hat t$ difference highlighted above leads to different scaling for  $(g^{(2)}_{WHq_Lq'_L} -g^{(4)}_{WHq_Lq'_L} )$  vs. $ g^{(3)}_{WHq_Lq'_L} $. For longitudinal $W$, the interference piece is significantly larger for the former combination 
\begin{align}
\label{eq:effW_dim8}
\int\limits^{\theta_{\text{max}}} d\theta^*\, 2{\,\rm Re}(A_{\rm SM}A^{(8)}_{3})_{W_L} \sim \frac{v^2\, \hat s^2}{\Lambda^4\, m^2_W}\, \quad\quad    \int\limits^{\theta_{\text{max}}} d\theta^*\, 2{\,\rm Re}(A_{\rm SM}A^{(8)}_{24})_{W_L} \sim \frac{v^2\, \hat s}{\Lambda^4}\,
\end{align}
where $A^{(8)}_{24}$ and $A^{(8)}_{3}$ stands for the pieces proportional to $(g^{(2)}_{WHq_Lq'_L} -g^{(4)}_{WHq_Lq'_L} )$ and $g^{(3)}_{WHq_Lq'_L}$, respectively. For SMEFT squared, this difference is even larger~\footnote{In scenarios where $g^{(3)}_{WHq_Lq_L}$ and $(g^{(2)}_{WHq_Lq'_L} -g^{(4)}_{WHq_Lq'_L} )$ are nonzero, we can get energy scaling between the extremes shown.}
\begin{align}
    \int\limits^{\theta_{\text{max}}} d\theta^*\,|A^{(8)}_{3}|^2_{W_L} \sim \frac{v^2\, \hat s^3}{\Lambda^8}\, \quad\quad    \int\limits^{\theta_{\text{max}}} d\theta^*\,|A^{(8)}_{24}|^2_{W_L} \sim \frac{v^2\, \hat s\, m^4_W}{\Lambda^8}\,
\end{align}

From the above expressions, we expect that the operators contributing to  $g^{(3)}_{WHq_Lq'_L} \propto c^{(4)}_{q^2H^2D^3}$ in terms of Table~\ref{tab:W_couplings} -- are the dominant dimension eight effect, at least among operators that generate $qV \to q'H$. Furthermore, restoring the Wilson coefficients to Eq.~\eqref{eq:effW_dim6} and Eq.~\eqref{eq:effW_dim8} taking their ratio, we again recover Eq.~\eqref{eq:coeffhierarchy}. As argued in Sec.~\ref{sec:dim8ops}, for fixed $\Lambda \sim $ few TeV and the dimension-six coefficient fixed to values consistent with LEP constraints, the natural suppression of $\mathcal O(1/\Lambda^4)$ relative to $\mathcal O(1/\Lambda^2)$ can be overcome by a hierarchy among coefficients and $c^{(4)}_{q^2H^2D^3}$ has a significant, if not dominant, effect.

The above analysis singles out $c^{(4)}_{q^2H^2D^3}$ as the dominant dimension eight operator, but it focuses on the longitudinal $W$ piece. While it's contribution to longitudinal $W$ is smaller, $(g^{(2)}_{WHq_Lq'_L}-g^{(4)}_{WHq_Lq'_L})$ has non-vanishing interference (in the effective $W$ limit) with the SM for transverse $W$. 
\begin{align}
    \int\limits^{\theta_{\text{max}}} d\theta^*\, 2{\,\rm Re}(A_{\rm SM}A^{(8)}_{24})_{W_T} \sim \frac{v^2\, \hat s}{\Lambda^4},\quad     \int\limits^{\theta_{\text{max}}} d\theta^*\, |A^{(8)}_{24}|^2_{W_T} \sim \frac{v^2\, \hat s\, m^4_W}{\Lambda^8}
\end{align}
The parametrics of this interference is weaker than the longitudinal case for $g^{(3)}_{WHq_Lq'_L}$, but this is offset somewhat by the larger transverse $W$ parton distribution function. The operators in $(g^{(2)}_{WHq_Lq'_L}-g^{(4)}_{WHq_Lq'_L})$ are $c^{(3)}_{q^2H^2D^3}$ and $c^{(3)}_{q^2WH^2D}$. The simple two-to-two argument does not tell us which effect -- transverse vs. longitudinal pdf or faster scaling with $\hat s$ -- dominates, and we need to rely on numerics.

Before we turn to simulation, there is one final feature of VBF present at $\mathcal O(1/\Lambda^4)$ that is absent at $\mathcal O(1/\Lambda^2)$ -- the contribution from five-particle (four quark plus one Higgs) vertices originating from the $H^2\psi^4$ class dimension eight operators shown in Table~\ref{tab:D8_nongeo_ops3}. These vertices enter VBF as contact terms (meaning no propagators) and can take on any of the helicity combinations allowed by the SM. The coupling factors that enter each helicity combination, as well as example $2\to 3$ amplitudes are given in Appendix~\ref{app:2to3amps}. We expect the largest contribution to come from operators with (LL)(LL) helicity structure (operators $Q_{q^4H^2}^{(1)}$ and $Q_{q^4H^2}^{(3)}$ in Table~\ref{tab:D8_nongeo_ops3}), as these can interfere with the largest SM helicity amplitude.  As $H^2\psi^4$ operators do not generate $qV \to q\,H$, we cannot study their energy enhancement relative to the SM using the techniques above, so we again need to use a numerical analysis.

\section{Resonant operators}
\label{sec:pol}

Having completed the amplitude analysis and clarified the energy enhancement arguments, we now numerically investigate the full \(2 \to 3\) process \(pp \to h jj\). For event simulation, we generated a Universal FeynRules Output (UFO) file using FeynRules~\cite{Alloul:2013bka}, which incorporates the operators listed in the Tables of Section~\ref{sec:dim8ops}. This model was then imported into \texttt{MadGraph5\_aMC@NLO}~\cite{Alwall:2014hca} for detailed numerical analysis.

We implement specific generation-level VBF selection criteria following Ref.~\cite{Araz:2020zyh}. We consider two benchmark cut points, one with minimum invariant mass of the final-state jets of $m_{j_1 j_2} > 480\,\text{GeV}$, along with a pseudorapidity separation of $\Delta\eta_{j_1 j_2} > 2.5$ between them. We further consider a more stringent invariant mass cut of $m_{j_1 j_2} > 600\,\text{GeV}$ with a pseudorapidity separation of $\Delta\eta_{j_1 j_2} > 3.0$. We further restrict the Higgs boson transverse momentum to the range $200\,\text{GeV} < p_T^H < 400\,\text{GeV}$.

We set the coefficients of the dimension-six operators to their values constrained by LEP measurements as derived in Ref.~\cite{Ellis:2020unq}. For the dimension-eight operators, we adopt the convention where all Wilson coefficients are set to unity, resulting in all operators being uniformly suppressed by a factor of $1/\Lambda^4$. We then vary the cutoff scale $\Lambda$ to investigate which dimension-eight operators can produce effects that are competitive with those of the dimension-six, LEP-constrained operators. Note that with the dimension-six operators fixed to their LEP-constrained values (the exact values we selected will be detailed below), varying the cutoff scale $\Lambda$ affects only the dimension-eight terms. In Section~\ref{sec:dim8ops}, we assumed that the dimension-six and dimension-eight operators share a common cutoff scale $\Lambda$. Applying that logic to the current context—where the dimension-six coefficients $c^{(6)}/\Lambda^2$ are fixed to LEP-compliant values—implies that as we vary $\Lambda$, we are implicitly adjusting the Wilson coefficients such that the combination $c^{(6)}/\Lambda^2$ remains constant.

While the scaling arguments presented in Section~\ref{sec:dim8ops} offer general scaling insights, they cannot precisely determine the appropriate choice of the cutoff scale \(\Lambda\) due to the complexities introduced by selection cuts. Therefore, we perform a numerical analysis by computing cross-sections and checking the validity of the EFT expansion, testing terms up to \((\text{dimension-eight})^2\), confirming that the cross-section of each operator turned on one at a time, at each order in the EFT, are $\mathcal{O}(1\%)$ of the previous order. Based on the selected ranges of \( m_{jj} \) and \( p_T^H \), we determine that the minimum allowed value of the cutoff scale \( \Lambda \) is approximately \( 1.2\,\text{TeV} \). The scale may seem low, but the suppressed interference effects require lower \(\Lambda\) values for the EFT contributions to become noticeable. This is expected from the \(\hat{s}\)-channel versus \(\hat{t}\)-channel argument discussed in Section~\ref{sec:dim8ops}. We further caution that for the most stringent cut we implement, $m_{j_1 j_2} > 600\,\text{GeV}$, the EFT validity is borderline for a subset of operators, so we recommend \( \Lambda>1.2\,\text{TeV} \) for higher $m_{j_1 j_2}$ cuts.

Consequently, we adopt \( \Lambda=1.2\,\text{TeV} \) for all subsequent plots and numerical analyses. As \( \Lambda \) increases, the contributions from dimension-eight operators diminish rapidly relative to those from dimension-six operators, scaling approximately as \( \big( \frac{1.2\,\text{TeV}}{\Lambda} \big)^4 \). This does not imply that larger values of \( \Lambda \) are uninteresting; rather, it indicates that higher energy scales \( E \) are necessary to observe significant effects from dimension-eight operators at these larger cutoff scales.

Table~\ref{tab:xsx6} presents the parton-level cross-sections to illustrate the effects of the dimension-six operators displayed in Table~\ref{tab:D6_biggest}, while Table~\ref{tab:xsx8} shows the effects of pure dimension-eight operators as given in Tables~\ref{tab:D8_nongeo_ops1},~\ref{tab:D8_nongeo_ops2} and~\ref{tab:D8_nongeo_ops3}. In the remainder of this section, we will refer to the operators by their corresponding Wilson coefficients. Moreover, we note that all simulations are performed up to the leading order in SM couplings. 

\begin{table}[ht]
\centering
\footnotesize
\renewcommand{\arraystretch}{1.4}
\begin{tabular}{|c|c|c|c|c|}
\hline
\textbf{Type} & $(480\,{\rm GeV},2.5)$  & SM Deviation (\%) & $(600\,{\rm GeV},3.0)$ &  SM Deviation (\%) \\
\hline
SM  & 0.1375(2)  & - & 0.1239(2) & -
 \\
\hline
$D=6$    & $0.1357(7)_{-0.0090}^{+0.0089}$  & $[-7.9, +5.2]$ & $0.1219(6)_{-0.0063}^{+0.0077}$ & $[-6.8, +4.5]$ 
 \\
\hline
$D=6+ (6\times 6)$  & $0.1355(7)_{-0.0077}^{+0.0087}$  & $[-7.1, +4.9]$ & $0.1221(6)_{-0.0065}^{+0.0080}$ & $[-6.8,+4.9]$  
\\
\hline
\end{tabular}
\caption{Cross-section of SM versus LEP constrained dimension six operators The kinematical cuts we choose: $200<p_T^H<400\,\rm GeV$ and $(m_{j_1j_2},\Delta\eta_{j_1j_2})$, as labeled in columns. The cross sections are for all SMEFT coefficients set to their central values and $95\%$ CL~\cite{Ellis:2018gqa, Ellis:2020unq}. Percentage deviations from the SM at $95\%$ CL are also given.   }
\label{tab:xsx6}
\end{table}

\begin{table}[htp!]
\centering
\renewcommand{\arraystretch}{1.4}
\footnotesize
\begin{tabular}{|c|c|c|c|c|}
\hline
\textbf{Type} & $(480\,{\rm GeV},2.5)$  & SM Deviation (\%) & $(600\,{\rm GeV},3.0)$ &  SM Deviation (\%) \\
\hline
SM  & 0.1375(2) & - & 0.1239(2) & - \\
\hline  
\multicolumn{5}{|c|}{Coefficients at $D=8$ }  \\
\hline
$c^{(1)}_{q^4H^2}$ & \textbf{0.1396(2)} & \textbf{+1.5} & \textbf{0.1261(2)} & \textbf{+1.8} \\
$c^{(2)}_{q^4H^2}$ & 0.1367(3) & 0.6 & 0.1234(2) & -0.4 \\
$c^{(3)}_{q^4H^2}$ & \textbf{0.1512(3)} & \textbf{+10.0} & \textbf{0.1359(2)} & \textbf{+9.7} \\
$c^{(1)}_{d^4H^2}$ & 0.1376(2) & +0.1 & 0.1240(2) & +0.1 \\
$c^{(1)}_{u^4H^2}$ & 0.1380(3) & +0.4 & 0.1250(2) & +0.9 \\
$c^{(1)}_{u^2d^2H^2}$ & 0.1374(3) & -0.1 & 0.1238(2) & -0.1 \\
$c^{(1)}_{q^2d^2H^2}$ & 0.1377(3) & +0.1 & 0.1222(3) & -1.4 \\
$c^{(2)}_{q^2d^2H^2}$ & 0.1370(3) & -0.4 & 0.1237(3) & -0.2 \\
$c^{(1)}_{q^2u^2H^2}$ & 0.1372(2) & -0.2 & 0.1239(3) & 0.0 \\
$c^{(2)}_{q^2u^2H^2}$ & 0.1385(2) & +0.7 & 0.1252(3) & +1.0 \\
$c^{(1)}_{q^2BH^2D}$ & 0.1374(3) & -0.1 & 0.1243(3) & +0.3 \\
$c^{(3)}_{q^2BH^2D}$ & 0.1374(3) & 0.0 & 0.1243(2) & +0.2 \\
$c^{(1)}_{q^2WH^2D}$ & 0.1375(2) & +0.2 & 0.1241(2) & +0.2 \\
$c^{(3)}_{q^2WH^2D}$ & \textbf{0.1408(3)} & \textbf{+2.4} & \textbf{0.1270(2)} & \textbf{+2.5} \\
$c^{(5)}_{q^2WH^2D}$ & 0.1372(3) & -0.2 & 0.1240(3) & +0.1 \\
$c^{(1)}_{u^2WH^2D}$ & 0.1381(2) & +0.4 & 0.1241(3) & +0.2 \\
$c^{(1)}_{u^2BH^2D}$ & 0.1375(3) & 0.0 & 0.1242(2) & +0.2 \\
$c^{(1)}_{d^2WH^2D}$ & 0.1373(3) & -0.1 & 0.1239(2) & 0.0 \\
$c^{(1)}_{d^2BH^2D}$ & 0.1375(3) & 0.0 & 0.1241(2) & +0.2 \\
$c^{(1)}_{q^2H^2D^3}$ & 0.1376(3) & +0.1 & 0.1240(2) & +0.1 \\
$c^{(2)}_{q^2H^2D^3}$ & 0.1372(3) & -0.2 & 0.1240(2) & +0.1 \\
$c^{(3)}_{q^2H^2D^3}$ & \textbf{0.1439(3)} & \textbf{+4.7} & \textbf{0.1299(2)} & \textbf{+4.8} \\
$c^{(4)}_{q^2H^2D^3}$ $(*)$ & \textbf{0.1419(3)} & \textbf{+3.2} & \textbf{0.1280(3)} & \textbf{+3.3} \\
$c^{(1)}_{u^2H^2D^3}$ & 0.1380(3) & +0.4 & 0.1244(3) & +0.4 \\
$c^{(1)}_{d^2H^2D^3}$ & 0.1371(2) & -0.3 & 0.1239(2) & 0.0  \\ 
\hline
\end{tabular}

\caption{Cross-section comparison of SM and dimension eight coefficients with $\Lambda = 1.2\phantom{t}{\rm TeV}$ and kinematical cuts: $200<p_T^H<400\,\rm GeV$ and $(m_{j_1j_2},\Delta\eta_{j_1j_2})$ labeled in columns. The percentage deviations are calculated relative to the SM cross-sections. The asterisk indicates that the operator breaks EFT validity for $\Lambda = 1.2\phantom{t}{\rm TeV}$ and one needs to set $\Lambda$ to a higher value in order to investigate it faithfully.}
\label{tab:xsx8}
\end{table}

Among the dimension-eight operators, we find that \(c_{q^4 H^2}^{(1,3)}\), \(c_{q^2 W H^2 D}^{(3)}\), and \(c_{q^2 H^2 D^3}^{(3)}\) have the largest contributions.  This observation is consistent with the fact that operators like \(c_{q^2 W H^2 D}^{(3)}, c_{q^2 H^2 D^3}^{(3)}\) appear in the coupling factors \(g^{(2)}_{WHq_L q'_L}, g^{(4)}_{WHq_L q'_L} \), which, as discussed in Section~\ref{sec:dim8ops}, generate \(qV_T \to q\,H\) terms that grow with energy. Contact terms like \(c_{q^4 H^2}^{(1,3)}\) are also important, and as previously argued, the largest effects come from contact terms with all left-handed chirality structures. These contact interactions involve direct couplings between quarks and the Higgs without intermediate propagators, leading to large effects at high energies. The differences we see lie in which contact operators lead to both $qqWH$ and $qqZH$ versus those that lead solely to $qqZH$. The latter effects are much smaller in comparison since $qqWH$ has a much larger impact on VBF in the SM part, which dimension eight operators interfere with. The remaining dimension eight operators have little effect. Lastly, we note that the large dimension eight contributions can result in  $\mathcal{O}(1-10\%)$ deviations from the SM, which is of the same order as the largest dimension six contributions given the same cuts and if the Wilson coefficients are set to LEP values.

The operator \(c_{q^2 H^2 D^3}^{(4)}\) also generates a significant impact, as expected from the previous section (in fact, we singled out \(c_{q^2 H^2 D^3}^{(4)}\) as the likely largest dimension eight effect). However, for this operator, choosing \(\Lambda = 1.2\,\text{TeV}\) leads to a breakdown of the EFT expansion. All other operators have squared dimension-eight contributions much smaller than their linear dimension-eight terms.  This behavior can be traced to the \(\hat{s}^4\) scaling in the squared SMEFT term and the fact that \(\hat{s} \gtrsim 1\,\text{TeV}\) under the cuts we are imposing. If we had chosen \(\Lambda \sim 3\,\text{TeV}\), the EFT expansion for this operator would be under control; however, the effects from all other dimension-eight operators would then be suppressed. Thus, we focus on the remaining operators which are valid at \(\Lambda = 1.2\,\text{TeV}\) given this is our choice and ignore \(c_{q^2 H^2 D^3}^{(4)}\) from now on as it requires a separate analysis at \(\Lambda > 3\,\text{TeV}\).

In summary, while the numerical values may vary depending on the kinematic cuts and experimental configurations, the central insights from our numerical study are: i) for VBF Higgs production, interference suppression means that the EFT cutoff can be relatively low, ii) dimension-eight operators can compete with dimension-six contributions under LEP constraints, and iii) especially when selecting lower cutoff scales, energy-dependent contributions can snowball and violate the perturbative regime, and further, depend strongly on the details of the operator structure. These considerations should guide future experimental analyses to capture these energy-enhanced higher-dimensional operators' impact fully.

\section{Observable distributions}\label{sec:distributions}

\begin{figure}[th!]
\centering
\includegraphics[width=0.49\textwidth]{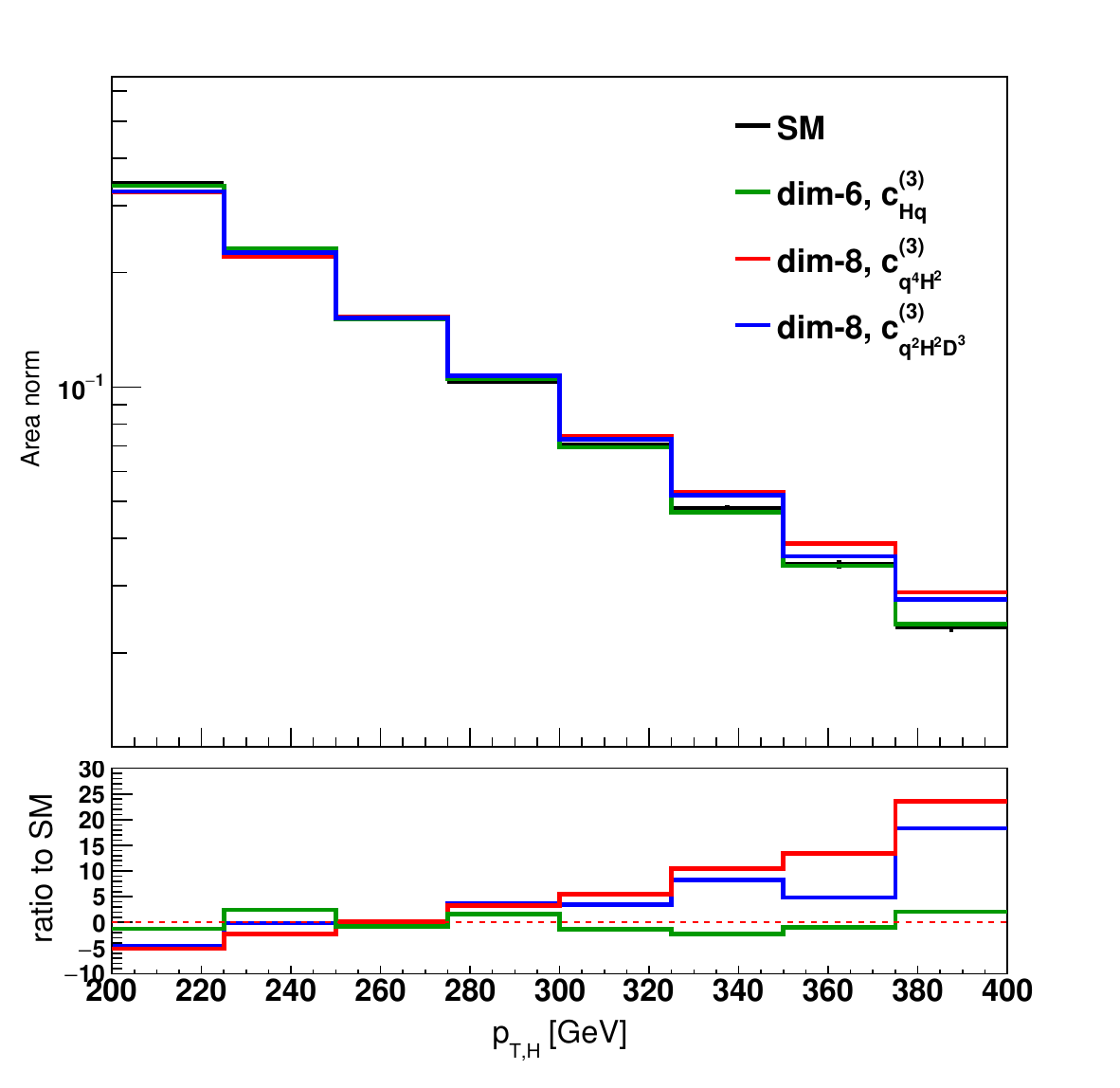}
\includegraphics[width=0.49\textwidth]{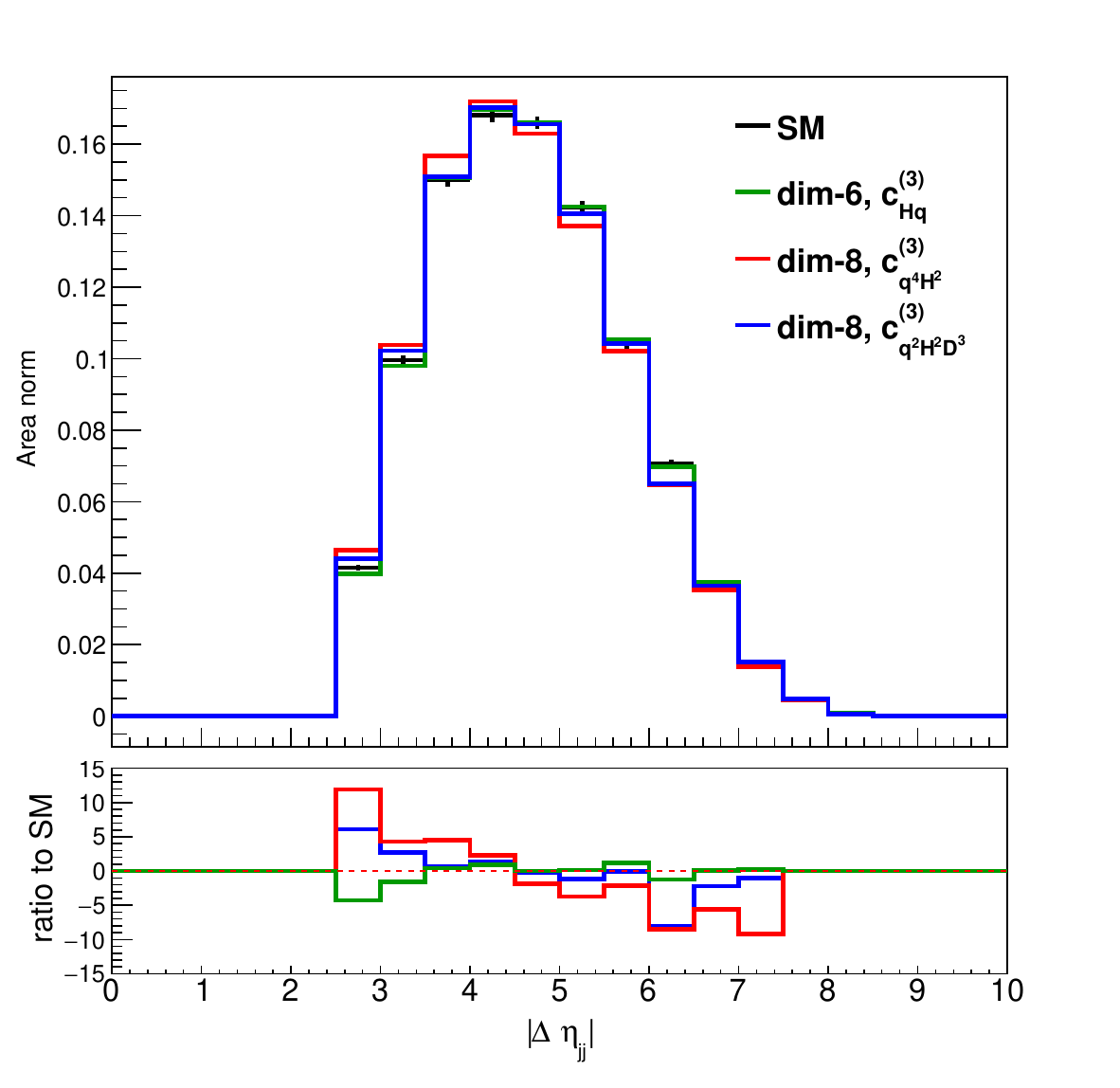}\\
\includegraphics[width=0.5\textwidth]{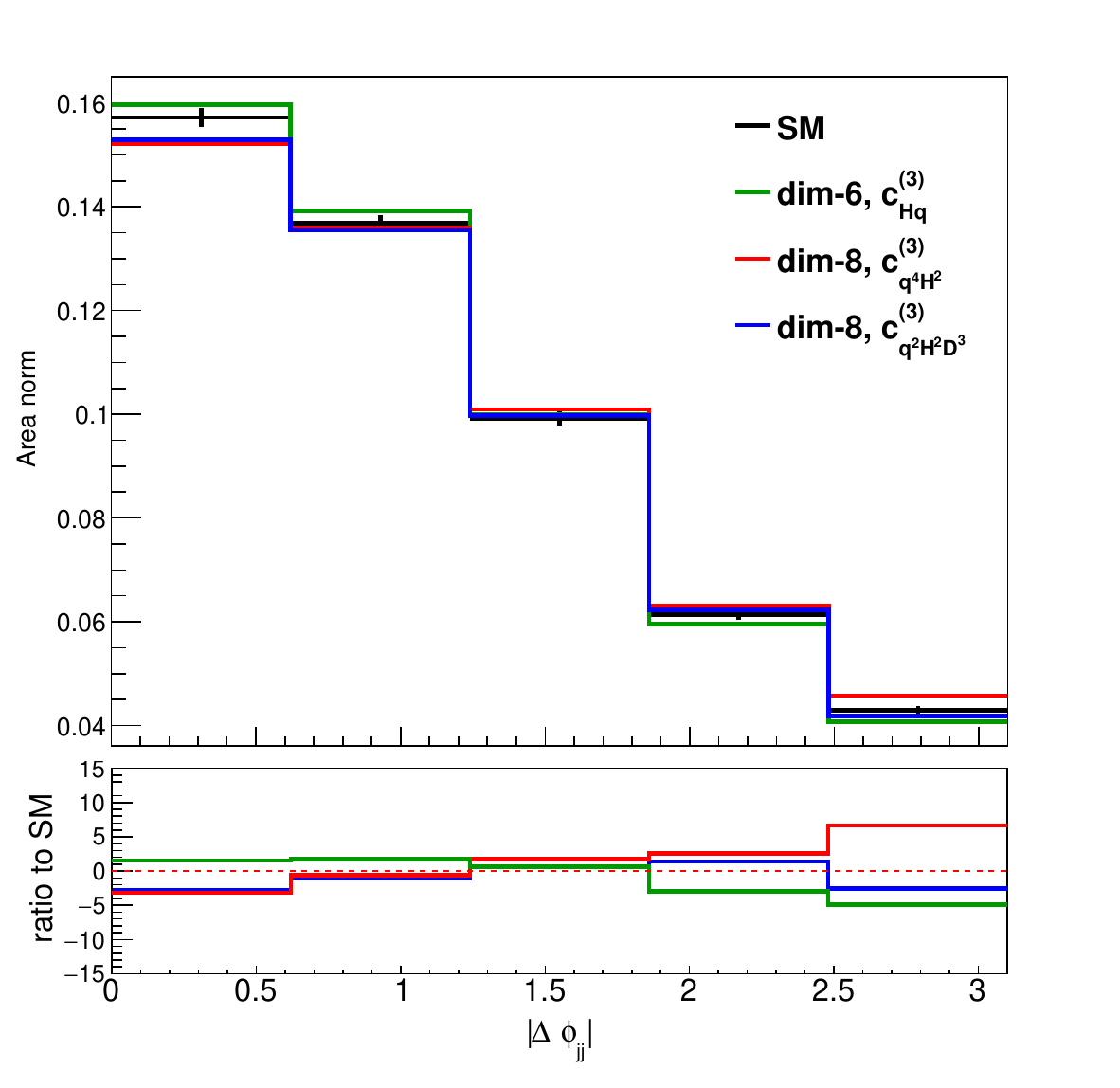}\\
\caption{ Top-left plot: Higgs $p_T$ in the range $200-400\, \text{GeV}$ in the SM and three different scenarios with one Wilson coefficient turned on at a time: $c^{(3)}_{Hq}, c^{(3)}_{q^4H^2}, c^{(3)}_{q^2H^2D^3}$ and taking $\Lambda = 1.2\, \text{TeV}$. $c^{(3)}_{Hq}$ is set to the upper 95 CL value consistent with EWPO and other data according to Ref.~\cite{Ellis:2020unq}, while $c^{(3)}_{q^4H^2}, c^{(3)}_{q^2H^2D^3}$ are set to $+1$. All curves are area normalized to emphasize shape differences -- differences in the overall rate in the different scenarios are tabulated in Table.~\ref{tab:xsx8}. The top-right and bottom plots show the $|\Delta \eta_{jj}|$ and $|\Delta \phi_{jj}|$ distributions under the same setup. }
\label{fig:higgspT}
\end{figure}

The differential distributions of key observables in VBF Higgs production~\cite{Plehn:2001nj,Plehn:2001qg,Buckley:2014ana}, such as \( p_T^H \), \( |\Delta \eta_{jj}| \), and \( \Delta \phi_{jj} \), are shown in Figure~\ref{fig:higgspT} to highlight the role of dimension eight operators within SMEFT. These results compare the SM predictions to those with SMEFT contributions up to \( \mathcal{O}(1/\Lambda^4) \), focusing on regions where dimension eight effects are most prominent.

In the top-left plot, the \( p_T^H \) distribution shows that, for the chosen parameters, dimension-eight operators influence the distribution more than dimension-six operators, particularly at high \( p_T^H \). This outcome aligns with expectations, as higher-dimensional operators exert more substantial effects at elevated energy scales. However, since both dimension-six squared and dimension-eight terms scale similarly, the small \( c_6 \) imposed by LEP constraints is what suppresses the dimension-six impact. Both $c^{(3)}_{q^4H^2}, c^{(3)}_{q^2H^2D^3}$ exhibit a similarly pronounced effect on the distribution, in particular, in the region $p_T^H>300\,\rm GeV$, compared to both the SM and the dimension-six operator \( c^{(3)}_{Hq} \). Note that the dimension eight operators enter by interfering with the SM, so the fact that we see more events at high $p_T^H$ can be traced to the sign choice for $c^{(3)}_{q^4H^2}, c^{(3)}_{q^2H^2D^3}$; had we picked $-1$ rather than $+1$, the magnitude of the deviation from the SM would be the same, but the interference would be destructive. Moreover, upon checking for Fig.~\ref{fig:higgspT} in the VBF context, in the \( p_T^H \) bin of 350–400~GeV, we find that the squared dimension‑eight contribution from \( c^{(3)}_{q^4H^2} \) is approximately 9\% of the SM interference, and for \( c^{(3)}_{q^2H^2D^3} \), the squared term is about 1\% of the interference. These findings confirm that the EFT expansion remains valid in this kinematic region.

In contrast, the angular distributions $\Delta \eta_{jj}$ and $\Delta \phi_{jj}$ shown in the top-right and bottom plots display more subtle differences between various SMEFT operators. These small changes in the distribution shapes could help distinguish which operator is responsible for any excess in high \( p_T^H \) events. Although the effects on angular variables are not as pronounced as in the \( p_T^H \) distribution, they provide complementary information that could assist in identifying specific operators contributing to deviations from the SM. In particular, the dimension six and eight coefficients $c^{(3)}_{Hq}$ and $c^{(3)}_{q^4H^2}$ have little effect on the angular distributions compared to the SM while $c^{(3)}_{q^2H^2D^3}$ shows a deviation. Thus, this observable can be used to disentangle dimension eight effects from $c^{(3)}_{q^2H^2D^3}$ and $c^{(3)}_{q^4H^2}$.

In terms of experimental implications, these findings suggest that analyses focusing solely on total rates may miss subtle yet significant effects arising from dimension-eight operators. Instead, differential measurements, particularly in the high \( p_T^H \) region, are needed for enhancing sensitivity to new physics. Additionally, careful consideration of kinematic cuts and binning strategies can optimize the detection of deviations caused by specific operators. Lastly, the correlations between different observables can provide further discriminatory power. For example, an excess in high \( p_T^H \) events accompanied by specific alterations in angular distributions could point toward certain operators over others. 


\section{Crossed process: Associated production $pp \to VH$}\label{sec:associated}


In the study of VBF, crossing an initial fermion line into a final one transforms the VBF topology into associated production, specifically $pp \to V(\bar{q}q)H$. The SMEFT operators that enhance energy effects in VBF are expected to influence the associated production of $VH$ similarly. The relevant Feynman diagrams are shown below in Fig.~\ref{fig:vh8-domin-diags}, where the shaded circles indicate an insertion of a SMEFT operator. Unlike VBF, this is not a \(t\)-channel process, so we do not expect differences in power counting from the amplitude squared level to the entire cross-section. In particular, we do not anticipate the SMEFT-SM interference to be suppressed, as was observed in the simple \(2 \to 2\) scattering case discussed in Section~\ref{sec:SMEFTcont}. 

\begin{figure}
    \centering
    \includegraphics[width=0.95\textwidth]{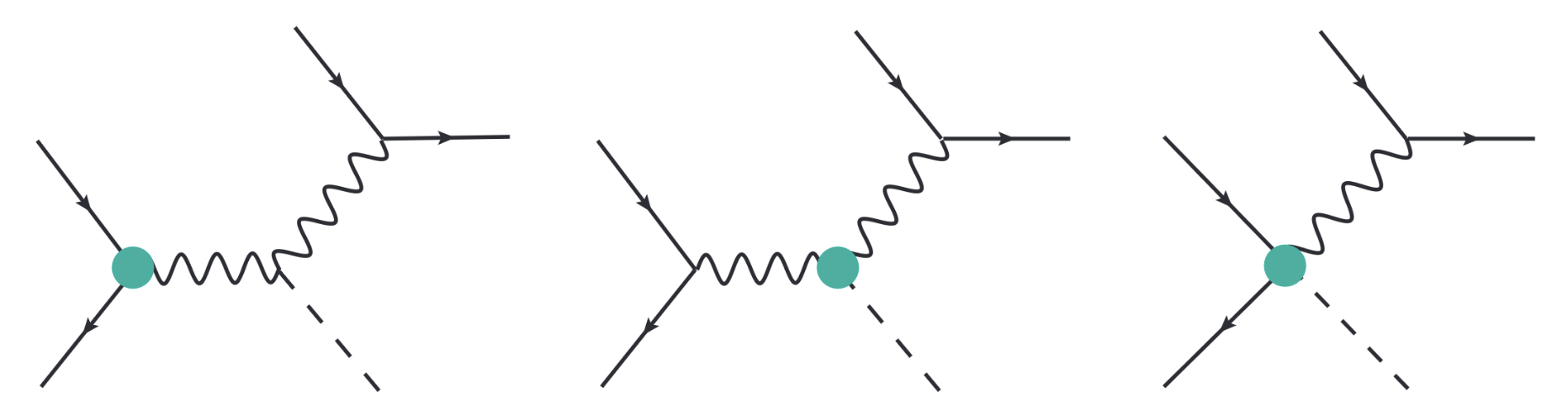}
    \caption{A subset of VH topologies with vector boson decay which illustrates the exact crossing symmetry between VH and VBF topologies. The green dots signify modified vertices from the EFT couplings.}
    \label{fig:vh8-domin-diags}
\end{figure}

Drawing on the potential for SMEFT-induced energy growth, several operators previously discussed in Section~\ref{sec:geobasis} were further analyzed in Ref.~\cite{Corbett:2023yhk}. That analysis, based on insights from Section~\ref{sec:dim8ops}, revealed that certain dimension-eight operators could significantly affect the process's kinematic tails, especially when dimension-six operators are constrained to comply with LEP measurements.
However, one specific class of operators -- with four fermions and two Higgses, $H^2\psi^4$ -- was not addressed in these studies, primarily because the focus of $VH$ studies typically centers on leptonic decays of $V$ rather than hadronic. Additionally, even if we were to look at hadronically decaying $V$, the expectation is that cuts designed to remove continuum (non-resonant) backgrounds would also suppress $H^2\psi^4$. This highlights how different analytical approaches in studying VBF versus VH can affect the observability of certain operators due to breaking crossing symmetry.

\begin{figure}[h!]
\centering
\includegraphics[width=0.6\textwidth]{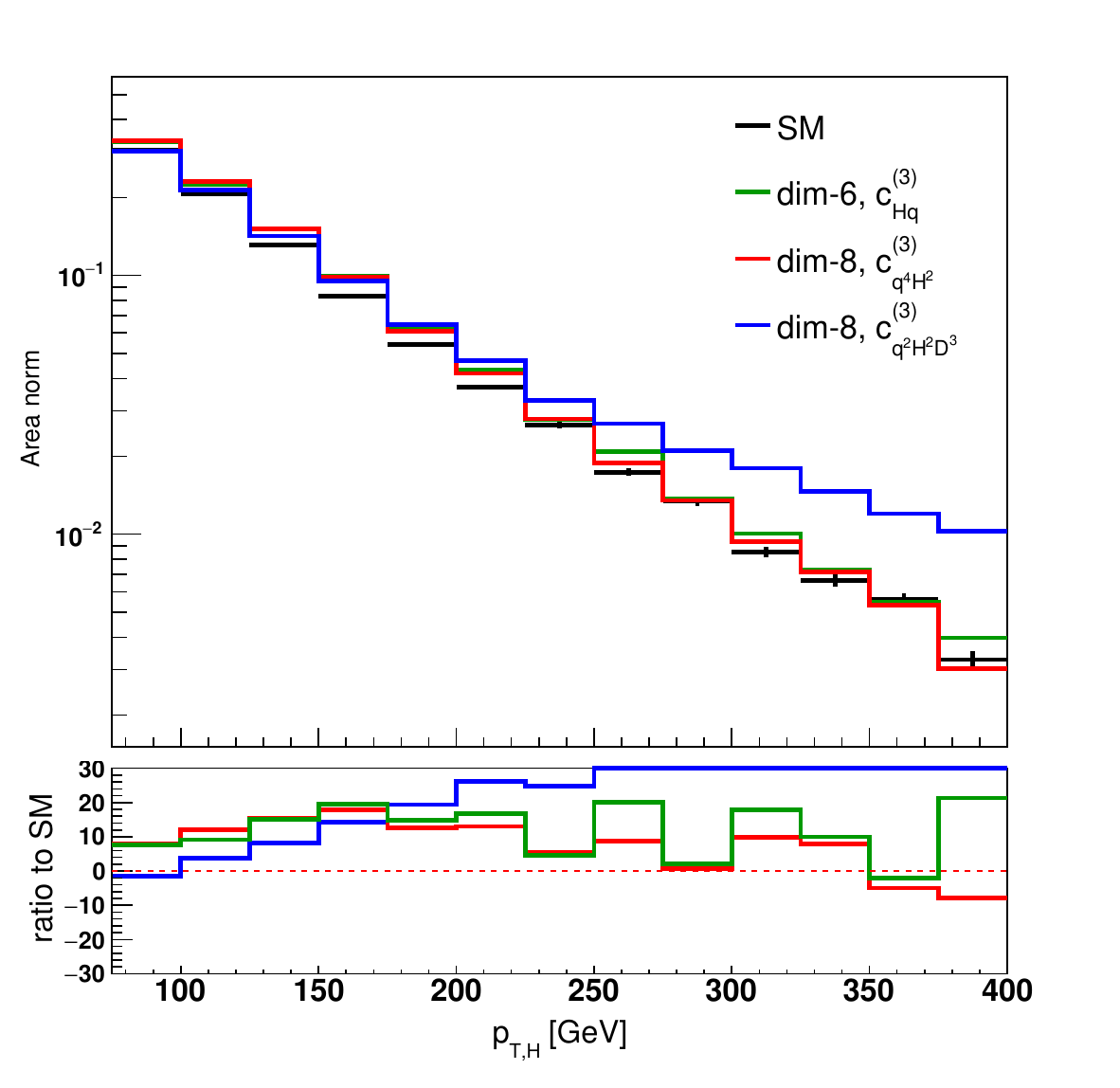}
\caption{Higgs $p_T$ in associated production in the SM and under the influence of several higher dimensional operators. As in Fig.~\ref{fig:higgspT}, $\Lambda = 1.2\, \text{TeV}$ for all SMEFT lines. $c^{(3)}_{Hq}$ is set to the upper 95 CL value from Ref.~\cite{Ellis:2020unq}, while the other coefficients are set to $+1$. In the ratio to SM sub-plot the blue line above $p_T^H = 250\, \text{GeV}$ is an artefact of the choice of plot range. }
\label{fig:higgsPTVH}
\end{figure}

To check this intuition, we simulated \( pp \to Z(\bar{q}q)H \) with cuts \( 75\, \text{GeV} \le p_{T,Z} \le 400\, \text{GeV} \) and \( 70 \le m_{jj} \le 110\, \text{GeV} \). These cuts were chosen to align with the STSX binning strategy.
The Higgs' resulting \( p_T \) distribution  turning on different operators is shown in Fig.~\ref{fig:higgsPTVH}.
We find that while operators \(c^{(3)}_{H^2 Q^4}  \) and \( c^{(3)}_{H^2 Q^2 D^3} \) have an impact on the Higgs \( p_T \) distribution that grows with \( p_T \). On the other hand, the dimension-six operator \( c^{(3)}_{Hq}\), set to its LEP-constrained value, has a much smaller impact relative to the SM. The lack of an effect from 'non-resonant' operators (meaning none of the quarks come from a \( W/Z \) decay) aligns with the conclusions from Ref.~\cite{Martin:2023tvi}.

We observe that the operator \( c^{(3)}_{H^2 Q^2 D^3} \) has a particularly large impact and approaches the boundary of the EFT's validity for the cuts we have applied. Combining this observation with the results from Fig.~\ref{fig:higgspT}, we conclude that a nonzero \( c^{(3)}_{H^2 Q^2 D^3} \) would lead to deviations in observables in both VBF and \( V H \) production that are correlated with the energy scale. Again, the exact magnitude of these deviations would depend on the specific cuts used in the analysis.

In contrast, the operator \( c^{(3)}_{H^2 Q^4} \) shows a negligible effect on \( V H \) production, as seen in Fig.~\ref{fig:higgsPTVH}. Even when considering scales \( \Lambda \) that push \( c^{(3)}_{H^2 Q^2 D^3} \) to the edge of EFT validity, \( c^{(3)}_{H^2 Q^4} \) remains suppressed due to the analysis cuts imposed. These cuts effectively break the crossing symmetry arguments, leading to deviations in VBF alone. The total cross-section remains largely unaffected by \( c^{(3)}_{H^2 Q^4} \) in \( V H \) production, reinforcing the notion that its impact is minimal in this context.

This comparison highlights the importance of analysis cuts and their influence on the observability of higher-dimensional operators. While \( c^{(3)}_{H^2 Q^2 D^3} \) affects both VBF and \( V H \) processes, \( c^{(3)}_{H^2 Q^4} \)'s impact is predominantly in VBF due to suppression in \( V H \) from the cuts designed to isolate resonant production. Therefore, operators like \( c^{(3)}_{H^2 Q^4} \) may require dedicated studies within VBF channels to be effectively constrained. Looking ahead, di-Higgs production via VBF could also be significantly affected by operators of the \( H^2 \psi^4 \) type. This makes VBF di-Higgs an instrumental setup for further studies, potentially allowing for combined analyses with VBF single Higgs data.

\section{Conclusions}\label{sec:conclude}

In this work, we have investigated the impact of dimension-eight operators in the SMEFT on VBF Higgs production. Our analysis focused on identifying operators that can introduce energy-enhanced contributions to the process, particularly in the context where LEP measurements constrain dimension-six operators. We find that operators introducing new Lorentz structures, different from those present in the SM, can lead to amplitudes that grow with energy. Specifically, dimension-eight operators involving derivatives or additional field content can introduce terms in the amplitude that scale with higher powers of the energy, potentially compensating for the suppression due to the higher dimensionality of the operator.

The difference between $s$-channel and $t$-channel processes was an essential additional consideration arising from our analysis. In VBF, a $t$-channel process, the SM amplitude exhibits a forward singularity as the momentum transfer squared \( \hat{t} \to 0 \). This leads to the cross-section being dominated by kinematic regions with small \( \hat{t} \) and large \( \hat{s} \), where \( \hat{s} \) is the center-of-mass energy squared of the subprocess. In such regions, the interference between the SM amplitude and SMEFT contributions can be suppressed due to the operators' different momentum dependencies and it being less singular than SM as $\hat{t}\to 0$. 

SMEFT operators contributing new Lorentz structures may not interfere effectively with the SM in the forward region, leading to suppressed interference terms. In contrast, $s$-channel processes, such as the crossed $VH$-production process, do not exhibit this suppression, as the interference between the SM and SMEFT amplitudes remains significant across the kinematic range. In order to assess this, we compared the impact of these operators in VBF with their effects in  (\( VH \)) production. We found that certain operators that significantly affect VBF may have negligible impact on \( VH \) production due to differences in the kinematic regimes and analysis cuts, which can suppress the contributions of non-resonant operators in \( VH \) channels.

Our analysis also revealed that the momentum dependence of the operators plays a crucial role in determining their contributions to the amplitude. Operators with derivatives acting on fields can introduce additional energy factors in the numerator, potentially enhancing their impact at high energies. However, in processes like VBF, where certain kinematic invariants dominate, the interplay between the momentum dependence of the numerator and the denominator can lead to non-trivial enhancements on the cross-section.  We then identified specific dimension-eight operators, such as those in the classes \( \psi^2 H^2 D^3 \) and \( H^2 \psi^4 \), that can lead to significant energy-enhanced contributions in VBF Higgs production. We found that these operators can induce observable deviations in distributions like the Higgs transverse momentum and jet angular separations, particularly in the high-energy tails of the distributions.

Importantly, we demonstrated that while dimension-six operators are often the focus of SMEFT analyses, dimension-eight operators can play a comparable or dominant role under certain conditions. This is especially true when experimental data constrain dimension-six operators, and their coefficients are small, allowing comparable dimension-eight contributions to become significant despite their naive suppression by higher powers of \( 1/\Lambda \). We further emphasized the need to consider the validity of the EFT expansion, mainly when dealing with energy-enhanced operators. We found that the EFT remains reliable for cutoff scales \( \Lambda \gtrsim 1.2\,\text{TeV} \) when applying standard VBF selection cuts for all operators except $c^{(4)}_{q^2H^2D^3}$. This underscores the importance of ensuring that the energy scales probed in the experiment are compatible with the EFT expansion to maintain the theory's predictive power.

\section*{Acknowledgements}

The work of AM is partially supported by the National Science Foundation under Grant Numbers PHY-2112540 and PHY-2412701, and the work of BA is supported by the Fermi National Accelerator Laboratory (Fermilab). Fermilab is managed by Fermi Research Alliance, LLC (FRA), acting under Contract No. DE--AC02--07CH11359. This work was performed in part at the Aspen Center for Physics, with support for BA by a grant from the Simons Foundation (1161654,Troyer).

\appendix

\section{Complete $2\to 3$ amplitudes}
\label{app:2to3amps}

\begin{table}[htp!]
\centering
\footnotesize
\renewcommand{\arraystretch}{1.4}
\begin{tabular}{ll}
\hline
\textbf{Coupling} & \textbf{Expression} \\
\hline \\ [-2.5ex]
\( g_{ZHu_Ru_R}^{(1)} \) & \( \displaystyle -\frac{e v}{c_w s_w}\, c_{Hu} \)\\  \\[-1.5ex] 
\( g_{ZHd_Rd_R}^{(1)} \) & \( \displaystyle -\frac{e v}{c_w s_w}\, c_{Hd} \) \\ \\[-1.5ex] 
\( g_{ZHu_Ru_R}^{(2)} \) & \( \displaystyle -\frac{v}{4 c_w s_w} \bigg( e\, c^{(1)}_{u^2H^2D^3} - 4 c_w s_w \big( c^{(1)}_{u^2BH^2D}\, s_w + 2 c^{(1)}_{u^2WH^2D}\, c_w \big) \bigg) \) \\
\\[-1.5ex]
\( g_{ZHd_Rd_R}^{(2)} \) & \( \displaystyle -\frac{v}{4 c_w s_w} \bigg( e\, c^{(1)}_{d^2H^2D^3} - 4 c_w s_w \big( c^{(1)}_{d^2BH^2D}\, s_w + 2 c^{(1)}_{d^2WH^2D}\, c_w \big) \bigg) \) \\
\\[-1.5ex]
\( g_{ZHu_Ru_R}^{(3)} \) & \( \displaystyle \frac{v}{4 c_w s_w} \bigg( 3 e\, c^{(1)}_{u^2H^2D^3} + 4 c_w s_w \big( c^{(1)}_{u^2BH^2D}\, s_w + 2 c^{(1)}_{u^2WH^2D}\, c_w \big) \bigg) \) \\
\\[-1.5ex]
\( g_{ZHd_Rd_R}^{(3)} \) & \( \displaystyle \frac{v}{4 c_w s_w} \bigg( 3 e\, c^{(1)}_{d^2H^2D^3} + 4 c_w s_w \big( c^{(1)}_{d^2BH^2D}\, s_w + 2 c^{(1)}_{d^2WH^2D}\, c_w \big) \bigg) \) \\
\\[-1.5ex]
\( g_{u_L d_L u_L d_L H} \) & \( \displaystyle 2 v \big( c^{(1)}_{q^4H^2} - c^{(3)}_{q^4H^2} + 2 c^{(3)}_{q^4H^2} \big) \) \\
\\[-1.5ex]
\( g_{u_L d_R u_L d_R H} \) & \( \displaystyle v \big( c^{(1)}_{q^2d^2H^2} - c^{(2)}_{q^2d^2H^2} \big) \) \\
\\[-1.5ex]
\( g_{u_R d_R u_R d_R H} \) & \( \displaystyle v\, c^{(1)}_{u^2d^2H^2} \) \\
\\[-1.5ex]
\( g_{u_L u_L u_L u_L H} \) & \( \displaystyle 2 v \big( c^{(1)}_{q^4H^2} - c^{(2)}_{q^4H^2} + c^{(3)}_{q^4H^2} \big) \) \\
\\[-1.5ex]
\( g_{u_R u_L u_R u_L H} \) & \( \displaystyle v \big( c^{(1)}_{q^2u^2H^2} - c^{(2)}_{q^2u^2H^2} \big) \) \\
\\[-1.5ex]
\( g_{u_R u_R u_R u_R H} \) & \( \displaystyle 2 v\, c^{(2)}_{u^4H^2} \) \\
\\[-1.5ex]
\( g_{d_L d_L d_L d_L H} \) & \( \displaystyle 2 v \big( c^{(1)}_{q^4H^2} + c^{(2)}_{q^4H^2} + c^{(3)}_{q^4H^2} \big) \) \\
\\[-1.5ex]
\( g_{d_R d_L d_R d_L H} \) & \( \displaystyle v \big( c^{(1)}_{q^2d^2H^2} + c^{(2)}_{q^2d^2H^2} \big) \) \\
\\[-1.5ex]
\( g_{d_R d_R d_R d_R H} \) & \( \displaystyle 2 v\, c^{(1)}_{d^4H^2} \) \\ [1.5ex]
\hline
\end{tabular}
\caption{Right-handed and five-point contact couplings with implicit powers of \(1/\Lambda\).}
\label{tab:rem_couplings_all}
\end{table}

We provide the full $2\to 3$ amplitudes for VBF. We do this to trace the trends of which operators are most energy enhanced at $2\to 2$ level to the full VBF process. Also, we would like to examine operators that do not contribute at the $2\to 2$ level but do enter in to $2\to 3$, so we need a unifying framework. We only show the $W$ exchange (sub) amplitude piece of $u_L(p_2) d_L(p_1) \to u_L(p_3) d_L(p_4) H(p_5)$. We pick this combination -- left handed (LH) quarks, $qq'$ initial state -- as it dominates the SM contribution, benefiting from larger LH quark couplings to the $Z$, the fact that both $W$ and $Z$ can serve as intermediate vector bosons, and larger quark parton distribution functions (versus antiquark, as would appear in a $q\bar q'$ initiated subprocess)

We break up the $2\to 3$ expression into three pieces, i.) pieces with SM topology, as shown in Figs.~\ref{fig:vbf-domin-diags} (a) and (b) which contain two vector boson propagators, ii.) SMEFT pieces containing a four-particle $qqVH$ vertex and a single vector boson propagator, see Fig.~\ref{fig:vbf-domin-diags} (c), and iii.) SMEFT pieces containing a five particle $qqqqH$ vertex and no propagators as shown in Fig.~\ref{fig:vbf-domin-diags} (d).  

The contributions from SM topologies in Fig.~\ref{fig:vbf-domin-diags} (a) and (b) are:
\begin{align}  
  \label{amps(a,b)}
{\cal A}^{(a,b)}(u_L d_L \to u_L d_L H) ={}& 
ig_{Wu_Ld_L}g_{Wu_Ld_L}^*\hat{p}_{24,W}\hat{p}_{13,W}\braket{43}[12] \bigg(  2g_{HWW}^{(1)}+ g_{HWW}^{(2)}R
\nonumber \\ & 
-g_{HWW}^{(2)}(s_{12}+s_{14}+s_{34}+s_{23})
  \bigg)  ,
\end{align}
where $R=\frac{\bra{4}\slashed{p}_H|2]\bra{3}\slashed{p}_H|1]}{\braket{43}[12]}$. Due to its dependence on $\slashed{p_H} \slashed{p_H}$, this term will scale like $m_H^2/p_{ij,W}$ (as opposed to something $\propto s_{12}$) upon interfering with the SM and after phase-space integration.

Next, the topology containing 4-point contact interactions shown in Fig.~\ref{fig:vbf-domin-diags} (c) has LH amplitude for $u_L d_L \to u_L d_L H$: 
\begin{align}  
  \label{amps(c)}
{\cal A}^{(c)}(u_L d_L \to u_L d_L H) &= 
ig_{Wu_Ld_L}g_{Wu_Ld_L}^*\hat{p}_{13,W}\bigg( 2g_{WHu_Ld_L}^{(1)}\braket{43}[12] +\braket{43}[12](g_{WHu_Ld_L}^{(2)}(s_{15}+s_{35})   \nonumber \\ &
+g_{WHu_Ld_L}^{(3)}(s_{35}-s_{15}))
+g_{WHu_Ld_L}^{(3)}\bra{4}\slashed{p}_H|2]\bra{3}\slashed{p}_H|1]\nonumber \\ &+g_{WHu_Ld_L}^{(4)}\bra{4}\slashed{p}_H|2]\bra{3}\slashed{p}_3-\slashed{p}_1|1]\bigg)
\nonumber \\ &
+ig_{Wu_Ld_L}g_{Wu_Ld_L}^*\hat{p}_{24,W}\bigg( 2g_{WHu_Ld_L}^{(1)}\braket{43}[12] +\braket{43}[12](g_{WHu_Ld_L}^{(2)}(s_{45}+s_{25})  \nonumber \\ &
+g_{WHu_Ld_L}^{(3)}(s_{25}-s_{45}))
+g_{WHu_Ld_L}^{(4)}\bra{4}\slashed{p}_H|2]\bra{3}\slashed{p}_H|1]\nonumber \\ &+g_{WHu_Ld_L}^{(3)}\bra{4}\slashed{p}_H|2]\bra{3}\slashed{p}_2-\slashed{p}_4|1]\bigg). 
\end{align}
Combining Eq.~\eqref{amps(a,b)} and~\eqref{amps(c)}, we can re-write to make energy-enhancement more explicit as,
\begin{align}  
  \label{amps_simp}
{\cal A}(u_L d_L \to u_L d_L H) &= 
2i\hat{p}_{24,W}\hat{p}_{13,W}g_{Wu_Ld_L}g_{Wu_Ld_L}^*\braket{43}[12]\bigg[\bigg(  g_{HWW}^{(1)} + \frac{1}{2}g_{HWW}^{(2)}R
\nonumber \\ & 
-\frac{1}{2}g_{HWW}^{(2)}(s_{12}+s_{14}+s_{34}+s_{23})
  \bigg)    \\&
+\frac{1}{\hat{p}_{24,W}}\bigg( g_{WHu_Ld_L}^{(1)} +\frac{1}{2}g_{WHu_Ld_L}^{(2)}(s_{15}+s_{35})+\frac{1}{2}g_{WHu_Ld_L}^{(3)}(s_{35}-s_{15})
\nonumber \\ &
+\frac{1}{2}g_{WHu_Ld_L}^{(4)}R+\frac{1}{2}g_{WHu_Ld_L}^{(3)}R_{31}\bigg)
\nonumber \\ &
+\frac{1}{\hat{p}_{13,W}} \bigg( g_{WHu_Ld_L}^{(1)}+\frac{1}{2}g_{WHu_Ld_L}^{(2)}(s_{45}+s_{25})+\frac{1}{2}g_{WHu_Ld_L}^{(3)}(s_{25}-s_{45})
\nonumber \\ &
+\frac{1}{2}g_{WHu_Ld_L}^{(4)}R+\frac{1}{2}g_{WHu_Ld_L}^{(3)}R_{24}\bigg)
\bigg] 
\end{align}
where $R_{ij}=\frac{\bra{4}\slashed{p}_H|2]\bra{3}\slashed{p}_i-\slashed{p}_j|1]}{\braket{43}[12]}$ and one can clearly see the enhancements manifest in $1/p_{ij,W}$ dependent terms as well as terms with $s_{kl}/p_{ij,W}$, $R/p_{ij,W}$, and $R_{ij}/p_{ij,W}$. 

Lastly, we examine the energy enhanced 5-particle contact term unique to dimension 8, \(q q \to q q H\) with $2\rightarrow 3$ amplitudes of the type,
\begin{align}  
{\cal A}( u_Ld_L \to u_Ld_LH) &= ig_{u_Ld_Lu_Ld_LH}  \langle 34 \rangle [12].
\label{eq:5ptamp}
\end{align}
The couplings $g_{\psi_1\psi_2\psi_3\psi_4H}$ are all $\propto v$ and start at $\mathcal{O} (\Lambda^{-4})$. Their scaling suggest that these amplitudes are unsuppressed in energy. The explicit forms of these couplings are provided in Table \ref{tab:rem_couplings_all}, where each term is expressed in terms of the operator coefficients \( c_i \) and electroweak parameters. This table further includes both dimension-6 right-handed couplings, dimension-8 right-handed couplings for reference.

The amplitudes for other helicity configurations (LLRR, RRLL, RRRR) have the same form as Eqn.~\eqref{amps(a,b)}-\eqref{eq:5ptamp} upon sending $\bra{i}\mathcal{O}|j]\rightarrow [i|\mathcal{O}\ket{j}$ for the right handed fermion currents and replacing all couplings with their right-handed versions.

\bibliographystyle{JHEP}
\bibliography{ref.bib}

\end{document}